\documentclass{ieeeaccess}

\usepackage{cite}
\usepackage{amsmath,amssymb,amsfonts}
\usepackage{graphicx}
\usepackage{textcomp}
\usepackage{multirow}
\usepackage{subcaption}
\usepackage{color,xcolor}
\usepackage{siunitx}
\usepackage{breqn}
\usepackage{textgreek}
\usepackage{multicol}
\usepackage{stfloats}
\usepackage{xcolor}
\usepackage{soul}
\usepackage{color,colortbl}

\def\startfigure{\vspace{0pt}\begin{figure}[ht]\center} 
\def\BibTeX{{\rm B\kern-.05em{\sc i\kern-.025em b}\kern-.08em
		T\kern-.1667em\lower.7ex\hbox{E}\kern-.125emX}}

\begin{document}
\history{Date of publication xxxx 00, 0000, date of current version xxxx 00, 0000.}
\doi{10.1109/ACCESS.2020.DOI}

\title{Understanding the Applicability of Terahertz Flow-guided Nano-Networks for Medical Applications}
\author{\uppercase{Sebastian Canovas-Carrasco\authorrefmark{1}, Rafael Asorey-Cacheda\authorrefmark{1}, Antonio-Javier Garcia-Sanchez\authorrefmark{1}, Joan Garcia-Haro\authorrefmark{1}, Krzysztof Wojcik\authorrefmark{2}, Pawel Kulakowski\authorrefmark{3}}}
\address[1]{Department of Information and Communication Technologies, Universidad Politécnica de Cartagena, 30202 Spain (e-mail: \{sebas.canovas, rafael.asorey, antoniojavier.garcia, joang.haro\}@upct.es)}
\address[2]{Jagiellonian University Medical College, Krakow, Poland (e-mail: krzysztof.wojcik@uj.edu.pl)}
\address[3]{Department of Telecommunications, AGH University of Science and Technology, 30-059 Krakow, Poland (e-mail: kulakowski@kt.agh.edu.pl)}

\tfootnote{This research was funded by the projects AIM, ref. TEC2016-76465-C2-1-R (AEI/FEDER, UE), ATENTO, ref. 20889/PI/18 (Fundación Séneca, Región de Murcia), and LIFE (Fondo SUPERA Covid-19 funded by Agencia Estatal Consejo Superior de Investigaciones Científicas CSIC, Universidades Españolas and Banco Santander). S. Canovas-Carrasco thanks the Spanish MECD for an FPU (ref. FPU16/03530) pre-doctoral fellowship. This work was also supported by the Polish Ministry of Science and Higher Education with the subvention funds of the Faculty of Computer Science, Electronics and Telecommunications of AGH University of Science and Technology.}


\corresp{Corresponding author: R. Asorey-Cacheda (e-mail: rafael.asorey@upct.es).}

\begin{abstract}
	Terahertz-based nano-networks are emerging as a groundbreaking technology able to play a decisive role in future medical applications owing to their ability to precisely quantify figures, such as the viral load in a patient or to predict sepsis shock or heart attacks before they occur. Due to the extremely limited size of the devices composing these nano-networks, the use of the Terahertz (THz) band has emerged as the enabling technology for their communication. However, the characteristics of the THz band, which strictly reduce the communication range inside the human body, together with the energy limitations of nano-nodes make the in-body deployment of nano-nodes a challenging task. To overcome these problems, we propose a novel in-body flow-guided nano-network architecture consisting of three different devices: i) nano-node, ii) nano-router, and iii) bio-sensor. As the performance of this type of nano-network has not been previously explored, a theoretical framework capturing all its particularities is derived to properly model its behavior and evaluate its feasibility in real medical applications. Employing this analytical model, a thorough sensitivity study of its key parameters is accomplished. Finally, we analyze the terahertz flow-guided nano-network design to satisfy the requirements of several medical applications of interest.
\end{abstract}
	

\titlepgskip=-15pt

\maketitle
\section{Introduction}
\label{sec:introduction}



Electromagnetic (EM) nano-communications have been attracting attention in recent years as a powerful technological advance enabling novel applications that could revolutionize different areas \cite{Akyildiz2010,Akyildiz2010a,Balasubramaniam2013,Lemic2019survey,Abadal2017,9164961}, with particular relevance to medicine \cite{Yang2019comprehensive,Stelzner2017,Ali2016,Chahibi2017,Felicetti2016,Kulakowski2017,Chude2017}. This emerging field deals with communication among size-constrained devices (known as nano-devices) forming nano-networks, and the connection between these nano-networks to the Internet paving the way for a new paradigm, the Internet of Nano-Things (IoNT) \cite{Akyildiz2010}. 

As the size of nano-devices can be scaled down to a few cubic micrometers [12], the integration of a communication module inside each nano-device is a challenging issue that has been thoroughly addressed in different works in the literature \cite{THzsource2020,Jornet2013,HOSSEININEJAD201734,Cabellos2015,LLATSER2012353,Jornet2014transceiver}. All these studies point to the miniaturization of the antenna embedded into each nano-device, which requires the use of THz frequencies to communicate, leading to the standardization of the THz band (0.1-10 THz) as the working frequency spectrum for EM nano-communications \cite{IEEE1906}. However, these high EM frequencies entail high path losses when the signal propagates throughout the medium, greatly reducing the communication range between nano-devices. This limitation becomes more restrictive when the physical medium has a high water content, as is the case of human body tissues (e.g., skin or fat) and fluids (e.g., blood), due to their high absorption coefficient at THz frequencies \cite{Yang2015,Piro2016,Abbasi2016,Canovas-Carrasco2018a}.

Even though these nano-devices present several technical limitations, they are able to perform simple tasks in scenarios where larger artificial devices would be excessively intrusive. This is of special relevance within the human body, where simple tasks (e.g., monitoring a biomarker and sending warnings) performed in vivo and in real time would significantly improve the effectiveness of disease diagnoses. Concretely, we focus on a nano-communication network composed of micrometer-sized devices (hereafter referred to as nano-nodes) located inside the human cardiovascular system to assist the immune system in detecting specific bacteria in blood, severe inflammatory response that may lead to sepsis, blood flow disorders, and ischemic heart diseases, as discussed later in this work. This type of network, where nano-nodes are in constant movement throughout the bloodstream, is known as a flow-guided nano-network \cite{Canovas-Carrasco2019}.

Flow-guided nano-networks are designed to overcome two important challenges that must be carefully investigated. First of all, the limited communication range inside the human body makes direct communication between nano-nodes and an external device unachievable \cite{Piro2016,Canovas-Carrasco2018a}. Secondly, the amount of energy that can be stored in a nano-node is scarce due to the constraints derived from downsizing the battery. To alleviate this restriction and provide a continuous source of energy, nano-nodes should use a piezoelectric nano-generator to harvest energy from the environment, as proposed in several works in the literature \cite{Jornet2012b,Canovas-Carrasco2018,Singh2018,Afsana2018,Demir2018,Donohoe2017,Mohrehkesh2014}. Still, the energy provided by the piezoelectric generator does not allow nano-nodes to be continuously working, so they need to alternate between idle (stand-by) and active cycles.

\begin{figure}[]
	\centering
	\includegraphics[width=0.42\textwidth]{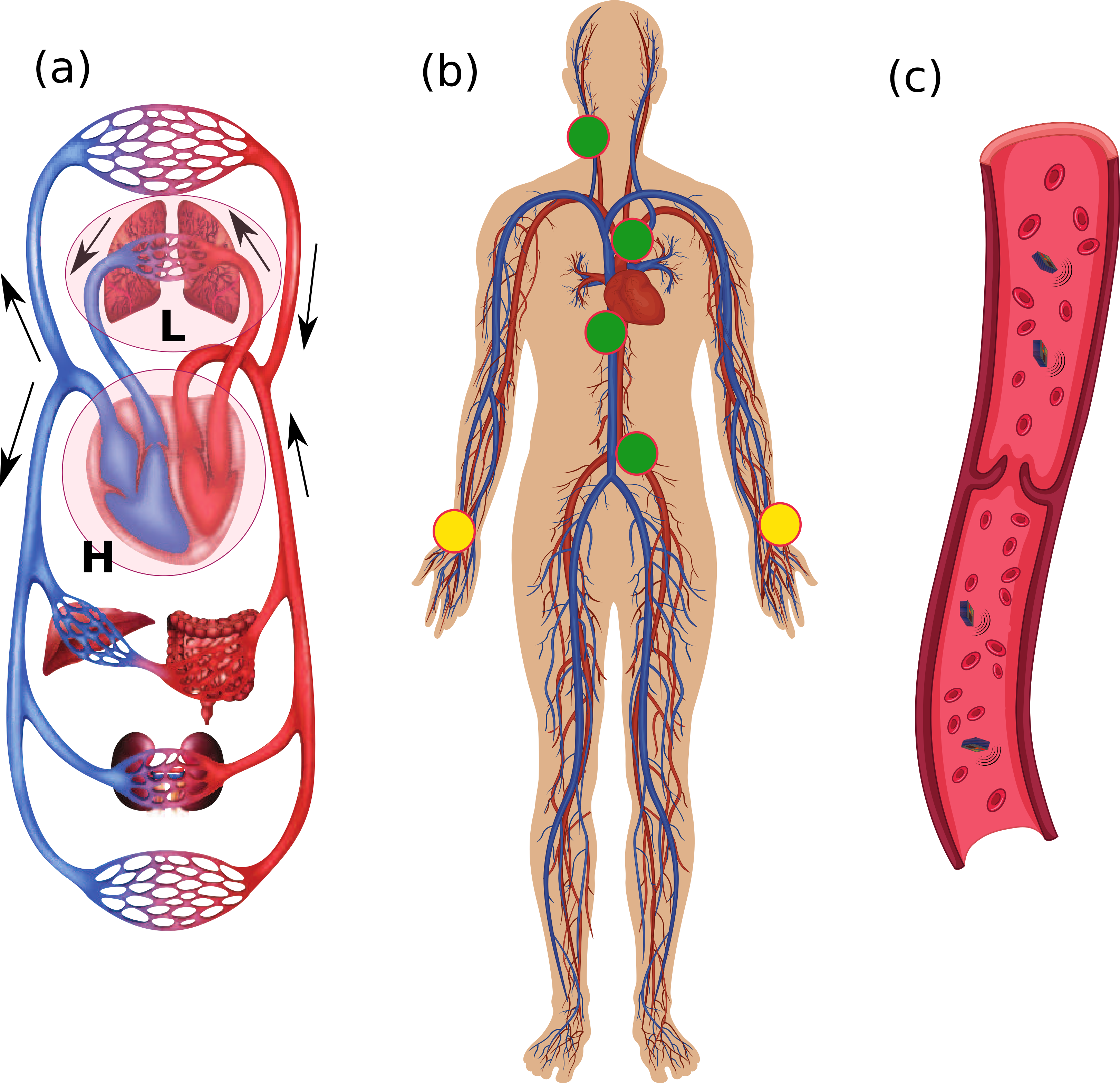}
	\caption{(a) The closed system of human blood circulation pumped by the heart (H) to the lungs (L) and other body organs; (b) bio-sensors (green circles) and nano-routers (yellow circles) indicated on the human vascular system; (c) mobile nano-nodes circulating along with the red blood cells in blood vessels. 
	}
	\label{fig:fig1}
\end{figure}

The human vascular system is a closed system, composed of many sub-circuits (see Fig. \ref{fig:fig1}a), so the nano-devices, when injected into it, travel within the bloodstream carried by the force of the pumping heart, reaching all the human tissues in subsequent cycles. Up to now, flow-guided nano-networks aimed at medical applications have been designed assuming that nano-nodes both sense the medium (i.e., blood) and transmit the information. However, this approach is not useful for certain applications where monitoring a critical part of the body is required. Thus, an additional sensing device positioned at the place of interest is necessary. With this objective in mind, we have conceived a new kind of flow-guided nano-network consisting of three different types of devices (depicted in Fig. \ref{fig:fig1}):

\begin{itemize}
	\item \textbf{Nano-node}. This is the smallest device, and it continuously flows through the blood circulatory system, being able to either receive (from a bio-sensor) or transmit data frames (to a nano-router). It is endowed with a THz communication module, in line with the proposals in \cite{Jornet2014transceiver,THzsource2020}, and a piezoelectric generator to recharge the battery periodically. Due to its extremely limited resources, a nano-node cannot always be operative, so it can only perform one task (transmission or reception), and only when its battery level is high enough.	
	\item \textbf{Nano-router}. A larger and less resource-constrained device that acts as a middle network tier receiving the data carried by nano-nodes and sending them to a macro-device able to connect the whole nano-network to a healthcare provider or, ultimately, to the Internet, following the paradigm of the IoNT \cite{Akyildiz2010,Balasubramaniam2013,AkyildizBio,Yu2018}. It is located on a superficial vein so that the distance between the nano-router and the macro-device (e.g., a wearable) is as short as possible [22].
	\item \textbf{Bio-sensor}. A medical device located in specific internal parts of the body sensing a given medical parameter or vital sign of interest. It is equipped with a communication module able to transmit the acquired information to the mobile nano-nodes circulating within the bloodstream.
\end{itemize}
As the characteristics of this type of flow-guided nano-network where nano-nodes act as information carriers are not similar to other nano-network topologies, a theoretical framework capturing all their particularities is necessary to properly model their behavior. For this purpose, in this paper we propose a general analytical model to mathematically evaluate different metrics of this novel kind of flow-guided nano-network. From this, we derive closed-form expressions for effective throughput, quality of delivery, and average data delay as functions of the different key parameters considered in the model. We also employ these expressions to analyze the impact of each one of these parameters on the nano-network performance. 
To summarize, the main contributions of this work are the following:
\begin{itemize}
	\item	We design and develop an analytical model capturing all the special characteristics of this novel THz flow-guided nano-network architecture based on Markov chains. This model jointly considers: (i) the energy balance in nano-nodes derived from the energy harvesting and consumption processes, and (ii) the probability of reception and transmission based on realistic values of the cardiovascular system and the limited communication range of nano-nodes due to communication in the THz band. We note that this model is general enough to be employed in different applications in which this nano-network could be used.
	\item	We provide insightful results about nano-network performance as a function of different parameters to accurately predict nano-network behavior in diverse scenarios. We pay special attention to the required number of nano-nodes in the network for less intrusiveness and an adequate network dimensioning for the different medical applications under study.
	\item	We validate the proposed model by simulating this type of flow-guided nano-network under different conditions. The results obtained from the simulations have been compared with those derived from the analytical expressions, showing that the mathematical framework accurately models the performance of this nano-network architecture.
	\item	We analyze the requirements and limitations of potential flow-guided nano-networks to be used in medical applications of early-stage detection of bacterial and viral infections, sepsis, heart attacks, and restenosis, employing realistic and accepted values for all the parameters employed.
\end{itemize}
The rest of the paper is summarized as follows. In Section \ref{sec:applications}, several potential medical applications where this kind of nano-network can play a critical role are identified. Section \ref{sec:analytical_model} rigorously defines the analytical model, considering all the characteristics of this new type of flow-guided nano-network. All the results are discussed in Section \ref{sec:results}, confirming the analytical model by means of simulations. The impact on network performance of the most relevant parameters is also analyzed and discussed. Finally, Section \ref{sec:conclusion} concludes the paper.

\section{Applications}
\label{sec:applications}

While a few years ago nano-devices in human blood were matters of theoretical concepts and early prototypes \cite{Douglas2012}, substantial progress has been made recently and numerous fully developed nano-bot systems have been presented. In \cite{Hu2013}, an effective application of nano-sponges cleaning the blood of toxins was reported. Later, the way a swarm of magnetotactic bacteria loaded with drugs and steered with a magnetic field could assist in tumor therapies was demonstrated \cite{Felfoul2016}. Also recently, nano-bots neutralizing pathogenic bacteria and toxins in blood have been presented \cite{Esteban2018}. In this section, we focus on some more challenging applications, where nano-devices should not only effectively perform their functions in the human vascular system, but also communicate with each other. Thus, examples of such applications related to bacteria and viral infections, sepsis, heart attacks, and restenosis, are discussed in detail.

\subsection{Bacterial blood infections}
While the presence of numerous bacteria types is quite common in a human body, their appearance in the vascular system is very dangerous, as they can then spread and infect multiple human organs. It is an especially serious problem in the case of hospital patients with many comorbidities, because their immune systems are already weakened. The ability to rapidly detect bacteria, like \textit{Pseudomonas}, \textit{Escherichia coli}, \textit{Acinetobacter}, \textit{Staphylococcus}, or \textit{Streptococcus} in cases of patients in a critical condition, might be a matter of life and death, as the early antibiotic administration is correlated with increased survival rate. Even a very low number of these bacteria in blood is dangerous. Thus, we propose a bio-sensor consisting of antibodies for each of these bacteria types. Antibodies (immunoglobulins) are protein molecules, about 10-30 nanometers in size, which recognize a epitope unique for different bacteria. Antibodies may signal the presence of the bacteria in many ways, but here, we focus on the technique described in \cite{Zhang2017}, where the antibody is attached to an electrode. After detecting the bacteria, the electrode resistance changes. Thus, assuming a simple micro-device is connected to the electrode, this device should be able to measure the resistance and start transmitting a warning immediately after detecting a change. A single sensor, no larger than a millimeter, can have antibodies detecting numerous important bacteria. Multiple sensors, such as these should probably be located close to important organs like lungs, the bladder, and kidneys, in order to identify the source of the bacteria. 

\subsection{Viral load monitoring}
The same technique proposed for bacteria detection \cite{Zhang2017} can be used for viral load, i.e., density monitoring \cite{Warkad2018}. Viruses are responsible for infectious diseases, but they differ from bacteria in a few important aspects. The main differences include: (a) the size, as viruses are significantly smaller than bacteria (ca. 10x); (b) different strategy of replication, as viruses use the host cell machinery to create their new copies. Viral infections can be divided into two groups. The first group are acute (severe, but short-lived) infections, like typical flu viruses, and also COVID-19. These viruses may not be present in the blood at all, thus the discussed monitoring system will not be able to identify them. The second group, however, are chronic (long-term) infections, in which case the viruses are detectable in blood. The most important infection of this group requiring viral load monitoring is the human immunodeficiency virus (HIV). This measurement is crucial to evaluate of therapy results, plan further treatment, and, most importantly, evaluate the risk of HIV transmission. The threshold for HIV viral load is 1500 copies/ml \cite{Marks2015}, and this is a cornerstone of the so-called U=U campaign (Undetectable Equals Untransmitable). Thus, this task will require a new function from the proposed system: not only the detection of certain particles but also the ability to quantitatively measure the concentration of studied elements. In the case of HIV, this is the number of copies per ml. 

\subsection{Sepsis}
Sepsis is a state of the human immune system when its reaction to infections may threaten its own healthy tissues and organs, finally causing death. Sepsis is signaled by a very high density of cytokines in the blood, in particular interleukin 6 (IL-6) molecules. IL-6 molecules are normally present in blood of a healthy individual as well, but in the case of sepsis, their density suddenly rises from about $1.5\cdot10^{10}$/ml (a healthy person) to $3.75 \cdot10^{11}$/ml and higher (for sepsis) \cite{Mardi2010}. They are produced by living cells and their lifetime may be modeled by an exponential distribution with an average of 45 minutes \cite{Castell1988}.  

Sepsis is usually detected via blood analysis in a laboratory. However, this is quite time consuming, as the blood must be taken from the patient, delivered to the lab and analyzed. Sepsis occurrence is very often detected too late, as it could be a matter of hours which determine whether a patient survives or not. The bio-sensor located inside the vascular system could check the patient's state constantly. However, as IL-6 density is quite high even in healthy people, sensors with antibodies cannot be used, as in the previous case with bacteria. Such a sensor would be immediately blocked with IL-6 cytokines, detecting and signaling their high density all the time. Instead, we propose using a special small tube, about 0.25 mm long, implanted inside of a human vein. At one end, the tube has a membrane that is semi-permeable, i.e., small molecules like IL-6 cytokines can flow through but living cells cannot (living cells are quite large, about 10-20 micrometers in diameter). The other walls of the tube are not permeable. So, the IL-6 can pass through the membrane and propagate inside the tube by diffusion, finally reaching the opposite end of the tube after about 15 minutes \cite{Milo2015}. At this end, we have a layer of antibodies matching IL-6 cytokines. When an antibody catches an IL-6, the electrical resistance of the connected electrode changes (as described in the previous subsection). The size of the semi-permeable membrane and the length of the tube are carefully chosen so that only a small number of IL-6 cytokines reach the antibodies for a healthy person. However, when sepsis appears, the number of IL-6 cytokines is much larger, so the nano-node connected to the electrode receives a much higher signal and should send a warning only in this case. 

With such a tube, the sensor with antibodies detecting IL-6 cytokines may work for a long time. As most of the IL-6 molecules brake when propagating through the tube, only a small amount of them reach the opposite end of it. After matching an IL-6, an antibody is not active during a certain period of time, but recovers after that and can work again. Moreover, as the membrane is semi-permeable, not letting living cells inside, the IL-6 cytokines cannot be produced inside the tube. A single tube in a vein is enough, because with sepsis, the density of IL-6 cytokines quickly rises in the entire vascular system. A second similar device could be considered as a back-up. 

\subsection{Heart attacks}
The discussed nano-network may also be critically helpful in cardiac issues, e.g., in the case of people at risk of heart attacks (myocardial infarctions). A marker which helps to detect the danger of a heart attack very early is a so called Heart-type Fatty Acid Binding protein (H-FABP). If the H-FABP density exceeds $1.8\cdot10^{11}$/ml, this is considered to be a serious warning that may suggest a myocardial infarction \cite{Reddy2016}. Here, we also propose a bio-sensor comprised of a layer of specific antibodies attached to an electrode, but the layer is not in a tube, instead, it is covered with a shutter, opening just for the time necessary to make a measurement. The shutter is normally closed, so the H-FABP present in the blood does not block the bio-sensor completely. The shutter opens periodically, once per e.g., 15 minutes, in order to measure H-FABP density. If the density is high, a warning signal is transmitted by the bio-sensor. It may also be activated externally in order to take the H-FABP measurement on demand. This procedure can be initiated if a person feels sick, e.g., has some clinical symptoms like chest pain, dyspnea (shortness of breath), or low tolerance to physical effort. Clinical symptoms correlated with high H-FABP levels are a clear indication of a myocardial infarction \cite{Reddy2016}. 

\subsection{Restenosis}
One more application is related to stenosis, i.e., the process of the narrowing of blood arteries, usually caused by atherosclerosis. A popular solution is to put a stent into the narrowed artery to keep it open. However, stents frequently get blocked after some time, as some thrombus (solidified blood) may gather on it and, in consequence, the artery becomes blocked again, which may cause a heart attack \cite{Buccheri2016}. In order to control the state of a stent, a micro-sensor can be mounted on it. The sensor on the stent would then periodically transmit signals to mobile nano-nodes flowing through the artery. When the process of restenosis becomes significant, which usually means the lumen of an artery is more than 90\% closed, i.e., the stent is covered with 1-1.5 millimeters of thrombus \cite{Buccheri2016}, the signals from the sensors on the stent will be much weaker. Thus, the network can detect this situation and dispatch a warning to the gateway and then to the external medical devices. About 20-30 sensors can be mounted on a single stent, in order to have clear information about its condition.

\section{Analytical Model}
\label{sec:analytical_model}

Considering the limitations implicitly derived from employing nano-devices, the analytical model for the proposed THz flow-guided nano-network is based on the following assumptions:

\begin{itemize}
	\item There are $n$ nano-nodes randomly distributed throughout the human vascular system, $n \geq 1, n \in N$, with a uniform distribution throughout the total volume ($V_{t}$). Therefore, the nano-node concentration per unit of volume can be considered as constant in the entire system. In addition, two types of larger (less resource-constrained) devices are placed in two different parts of the system: (i) the nano-router, which is in charge of receiving and gathering all the information transmitted by nano-nodes (encapsulated in data frames), and (ii) the bio-sensor, which is continuously transmitting the collected data from the area of interest to the nano-nodes.

	\item Nano-nodes constantly move through the cardiovascular system, which can be modeled as a branched closed circuit (i.e., each branch would be equivalent to a vein or an artery). The nano-router is placed in one of these branches to receive the data frames previously taken in another branch, where the bio-sensor is located. Although the time to complete a round depends on the branch through which a nano-node flows, on average, nano-nodes take $T$ seconds to complete a round through the circuit. 
	
	\item As the vascular system is divided into different subcircuits, the blood only flows through one of them in each round before returning to the lungs. This implies the inability of a nano-node to reach both bio-sensor and nano-router in just one round if they are located in separate subcircuits. Therefore, to prevent the overestimation of the probability of transmitting a previously received frame, we assume that the full data transmission from the bio-sensor to the nano-router requires at least 2 rounds, one round to receive the frame from the bio-sensor and the following one to transmit it to the nano-router.
	
	\item Due to the motion of nano-nodes within the bloodstream, the integrated nano-generator periodically charges the battery every $1/f$ seconds, with $f$ as the frequency of charge. Thus, nano-nodes alternate idle and active cycles. Due to battery and memory constraints, we consider the worst-case scenario, that is, that nano-nodes can only transmit or receive one data frame per active cycle. During idle cycles, nano-nodes just keep the frame stored in memory.
	
	\startfigure
	\includegraphics[width=0.45\textwidth]{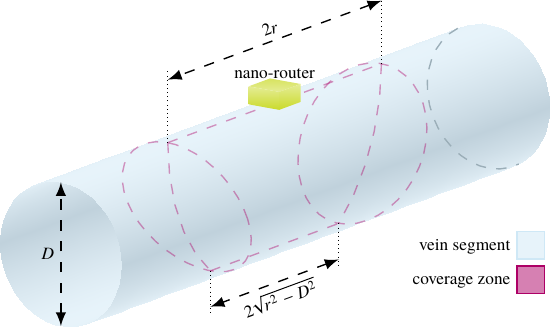}
	\caption{Coverage volume model.}\label{fig:fig_intersection}
	\end{figure}

	\item As the communication distance of the nano-node/nano-router link is greatly limited by the high path loss of the THz band experienced in blood\cite{Yang2015,Canovas-Carrasco2018a}, we define $V_{cv}$ as the volume in which nano-nodes are able to send a frame to the nano-router, that is, the coverage volume. This volume is defined by two parameters: (i) the communication range ($r$) and (ii) the diameter of the vein on which the nano-router is placed ($D$). Assuming an isotropic THz antenna is used \cite{Jornet2014}, the communication range of nano-nodes can be considered constant across the three dimensions of space. We remark that the model can be adapted to any type of transceiver just by tuning $V_{cv}$ as needed. Thus, generally speaking, the volume in which nano-nodes could communicate with the nano-router would be a sphere with a radius $r$ and the nano-router at its center. However, all the nano-nodes are confined within a vein, so the real volume $V_{cv}$ in which nano-nodes can indeed carry out a successful transmission is determined by the intersection between the already mentioned sphere and a cylinder with diameter $D$, as shown in Fig. \ref{fig:fig_intersection}. The expression required to analytically calculate $V_{cv}$ using Cartesian coordinates is defined by:
	\begin{equation}\label{eq:intersection_integral}
	V_{cv} = \int_{\zeta_{min}(x,y)}^{\zeta_{max}(x,y)} \int_{\gamma_{min}(x)}^{\gamma_{max}(x)} \int_{\chi_{min}}^{\chi_{max}} dx dy dz,
	\end{equation}
	\noindent where:
	\begin{equation}
		\chi_{min} = \begin{cases}
		-\frac{D}{2}  & \text{if $r^2 \geq \frac{D^2}{2}$}\\
		-r\sqrt{1-\frac{r^{2}}{D^2}} &  \text{if $r^2 < \frac{D^2}{2}$}
		\end{cases}
	\end{equation}
	\begin{equation}
		\chi_{max} = \begin{cases}
		\frac{D}{2}  & \text{if $r^2 \geq \frac{D^2}{2}$}\\
		r\sqrt{1-\frac{r^{2}}{D^2}} &  \text{if $r^2 < \frac{D^2}{2}$}
	\end{cases}
	\end{equation}
	\begin{equation}
	\!\!\gamma_{min}(x)\! = \!\max\!\left\{\frac{-D\!-\!\sqrt{D^2\! -\! 4 x^2}}{2},\! -\sqrt{r^2 - x^2}\right\}
	\end{equation}
	\begin{equation}
		\gamma_{max}(x) = \frac{-D+\sqrt{D^2 - 4 x^2}}{2}		
	\end{equation}
	\begin{equation}
		\zeta_{min}(x,y) = -\sqrt{r^2-x^2-y^2}		
	\end{equation}
	\begin{equation}
		\zeta_{max}(x,y) = \sqrt{r^2-x^2-y^2}		
	\end{equation}
%
%
%
	
	
	\item Depending on the part of the circulatory system through which nano-nodes flow, their velocity is variable. To model this velocity variation, we consider $v$ as the velocity when passing through $V_{cv}$. This velocity influences the time nano-nodes are within the coverage volume, thus affecting the probability of successfully transmitting a frame when crossing $V_{cv}$. In contrast, the velocity variation outside this zone does not impact on the metrics considered in this model, since it is characterized as an average round time ($T$), a known value studied in the literature (60 seconds) \cite{circulatorysystem}. Using this approach, we are able to develop, a priori, a model for any circular fluid system just by knowing the average round time, thus not needing the instantaneous velocity of nano-nodes when flowing through every part of the circuit, which can be really challenging in complex scenarios (e.g., the human cardiovascular system). 
	
	\item Without limiting the generality, we assume that nano-nodes cannot perform more than one transmission when crossing $V_{cv}$ since the time between transmissions is longer than the time to cross $V_{cv}$. As shown in Fig. \ref{fig:fig_intersection}, this assumption can be expressed mathematically as $2r/v \ll 1/f$. Therefore, nano-nodes can only perform one transmission per round.
	
	\item A successful transmission can only be achieved when a nano-node starts and ends the transmission of a data frame within $V_{cv}$. Thus, with $t_f$ as the time to transmit a complete frame, we define the transmission volume ($V_{tx}$) as:
	\begin{equation}
	\label{eq:Vtx}
	V_{tx} = \int_{\zeta_{min}(x,y)}^{\zeta_{max,tx}(x,y)} \!\!\!\! \int_{\gamma_{min,tx}(x)}^{\gamma_{max}(x)} \! \int_{\chi_{min,tx}}^{\chi_{max,tx}} dx dy dz,
	\end{equation}
	\noindent where $\chi_{min,tx}$, $\chi_{max,tx}$, $\gamma_{min,tx}$, and $\zeta_{max,tx}$ correspond to expressions \eqref{eq:mintx} to \eqref{eq:maxtx}.
	\begin{figure*}[b!]
		\vspace*{-\baselineskip}
		\begin{dmath}\label{eq:mintx}
			\chi_{min,tx} = \begin{cases}
				- \frac{D}{2}  & \text{if $r^2 \geq \frac{D^2}{2} + \left(\frac{v t_f}{2} \right)^2$}\\
				- \sqrt{\left(r^2 - \left(\frac{v t_f}{2}\right)^2 \right)\left(1 - \frac{1}{D^2}\left(r^2 - \left(\frac{v t_f}{2} \right)^2 \right) \right)} &  \text{if $r^2 < \frac{D^2}{2} + \left(\frac{v t_f}{2} \right)^2$}
			\end{cases}
		\end{dmath}
		\begin{equation}
			\chi_{max,tx} = \begin{cases}
				\frac{D}{2}  & \text{if $r^2 \geq \frac{D^2}{2} + \left(\frac{v t_f}{2} \right)^2$}\\
				\sqrt{\left(r^2 - \left(\frac{v t_f}{2}\right)^2 \right)\left(1 - \frac{1}{D^2}\left(r^2 - \left(\frac{v t_f}{2} \right)^2 \right) \right)} &  \text{if $r^2 < \frac{D^2}{2} + \left(\frac{v t_f}{2} \right)^2$}
			\end{cases}
		\end{equation}
		\begin{equation}
			\gamma_{min,tx}(x) = \max\left\{\frac{-D-\sqrt{D^2 - 4 x^2}}{2}, -\sqrt{r^2 - \left(\frac{v t_f}{2} \right)^2 - x^2} \right\}
		\end{equation}
		\begin{equation}\label{eq:maxtx}
			\zeta_{max,tx}(x,y) = \sqrt{r^2-x^2-y^2} - v t_f	
		\end{equation}
	\end{figure*}

	\item The probability of a nano-node being in the zone where a frame can be completely transmitted ($p_{tx}$) is modeled as the ratio between the transmission volume ($V_{tx}$) and the total fluid volume ($V_t$), that is:
	\begin{equation}  
	\label{eq:ptx}
	p_{tx}=\frac{V_{tx}}{V_t} 
	\end{equation}

	\item Even though the TS-OOK modulation presents a really low probability of interference between pulses (due to the extremely short pulse duration), it is required perfect multiplexing at reception to distinguish pulses from different transmitters and successfully demodulate more than one frame simultaneously. Thus, in order to keep the probability of receiving a flawed frame as low as possible and not to overestimate the performance of the nano-network, we assume that a collision occurs when the nano-router simultaneously receives two transmissions from two or more different nano-nodes. Hence, if a nano-node is performing a transmission within $V_{tx}$, a second nano-node can collide in two different cases: (i) it starts the transmission outside $V_{cv}$ but enters while the transmission is still active and (ii) it starts a transmission in any part of $V_{cv}$. So, there is a collision volume ($V_{cx}$) larger than $V_{cv}$ defined by: 
	\begin{equation}
	\label{eq:pcx}
	V_{cx} = \int_{\zeta_{min,cx}(x,y)}^{\zeta_{max}(x,y)} \int_{\gamma_{min}(x)}^{\gamma_{max}(x)} \int_{\chi_{min}}^{\chi_{max}} dx dy dz,
	\end{equation}
	\noindent where:
	\begin{equation}
	\zeta_{min,cx}(x,y) = -\sqrt{r^2-x^2-y^2}	- v t_f
	\end{equation}
	
	\item Similarly to the definition of $p_{tx}$, the probability of a nano-node being in the collision zone ($p_{cx}$) is modeled as:
	\begin{equation}
	p_{cx}=\frac{V_{cx}}{V_t} 
	\end{equation}
		
	\item As the bio-sensor is a larger device strategically located within a vein or an artery, covering its whole section, we consider that the probability of a nano-node passing through it in each round relies on the fraction of the total blood flow ($\eta$) that circulates through the vein/artery in which the bio-sensor is placed. In addition, the energy required by a nano-node to listen to the medium and receive a bit is much lower than to transmit it (approx. 10 times lower \cite{Jornet2012b}) and could even be passively detected without a preamplification stage \cite{Canovas-Carrasco2019}. As the presence of a bio-sensor is detected only once per round, the required energy for reception has a negligible impact on the total energy consumption of a nano-node. Therefore, we assume that nano-nodes are able to receive a frame if passing through the bio-sensor vein. Consequently, the probability of a nano-node receiving a frame in each round is: $p_{rx} = \eta$. 
%
%
\end{itemize}

For the sake of illustration, Fig. \ref{fig:router_nointerference} shows a nano-node flowing through the bloodstream communicating with a nano-router attached to a vein. As the transmission of a frame is performed every $1/f$, many other nano-nodes remain discharged when passing through $V_{cv}$ and are not able to successfully send data to the nano-router. Likewise, other nano-nodes are transmitting outside the coverage zone, so they do not cause any collision.


Based on these assumptions and considering that the frame is stored during two rounds (the minimum storage time to carry out a transmission from the bio-sensor to the nano-router), the throughput achieved by the proposed nano-network can be modeled by this expression:
\begin{equation}
\begin{split}
\label{eq:throughput}
Th & = n f p_{tx} p_{rx} \left(1-p_{cx}p_{rx}\right)^{n-1} = \\ 
  	& = n f \frac{V_{tx}}{V_t}\eta \left(1-\frac{V_{cx}}{V_t}\eta\right)^{n-1}
\end{split}
\end{equation}

This metric measures the frames per second that can successfully reach the nano-router. In order to achieve a correct transmission, nano-nodes must receive an updated frame, store it in their memories, and then transmit it to the nano-router without collisions. As can be seen, this model is generic enough to be employed in different applications, where any of the considered parameters may vary its value.		

\Figure[t!](topskip=0pt, botskip=0pt, midskip=0pt){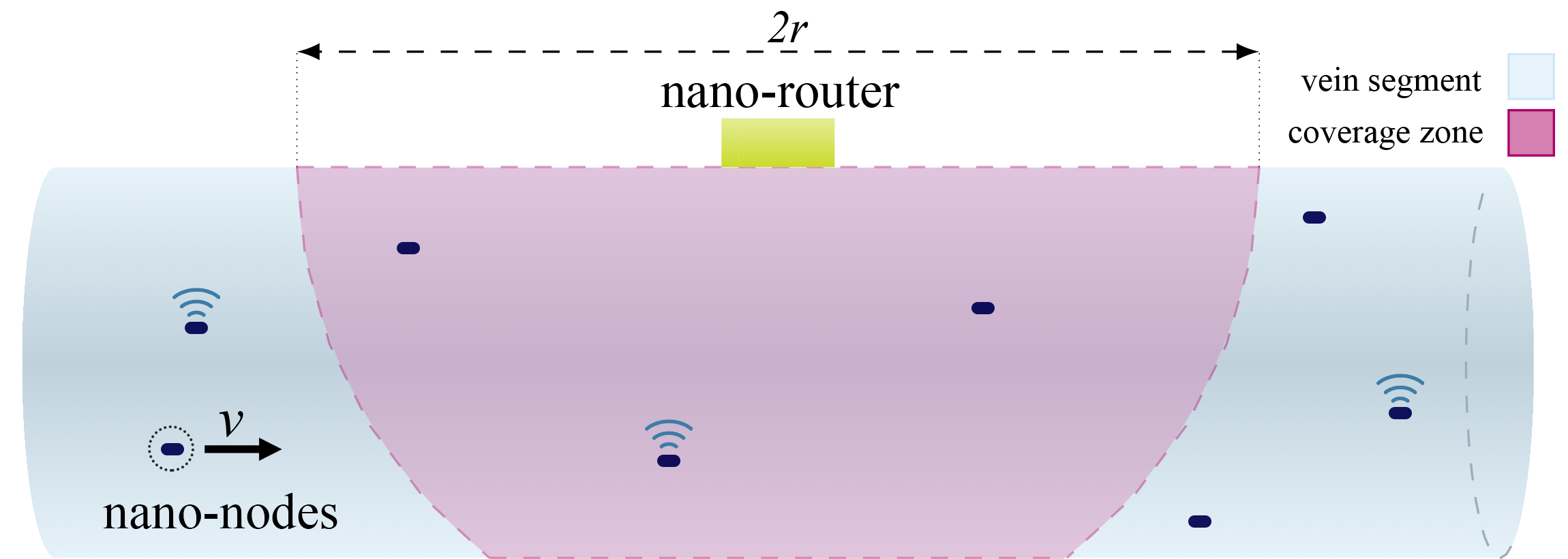}
{Nano-node communicating with the nano-router attached to a vein without collisions.\label{fig:router_nointerference}}

\begin{table*}[]
	\centering
	\renewcommand{\arraystretch}{1.2}
	\caption{Abbreviations employed throughout Section \ref{sec:analytical_model}.}
	\begin{tabular}{c c} 
		\hline
		Parameter & Description \\
		\hline
		$n$ & Number of nano-nodes\\		
		$T$ & Time to complete a round through the system\\
		$V_t$ & Total fluid volume\\
		$V_{cv}$ & Coverage volume\\
		$V_{tx}$ & Transmission volume\\
		$V_{cx}$ & Collision volume\\
		$D$ & Diameter of the vein where the nano-router is placed\\
		$r$ & Communication range of nano-devices\\
		$v$ & Nano-node velocity when passing through $V_{cv}$\\
		$t_f$ & Time to transmit a complete frame\\
		$1/f$ & Time between active cycles\\
		$\eta$ & Fraction of fluid circulating through the bio-sensor\\
		$p_{tx}$ & Probability of a nano-node being in the transmission volume\\
		$p_{cx}$ & Probability of a nano-node being in the collision volume\\
		$p_{rx}$ & Probability of receiving a frame per round\\
		$p_{s}$ & Probability of successfully transmitting a frame per active cycle\\
		$p_{s,rnd}$ & Probability of successfully transmitting a frame per round\\
		$k$ & Number of rounds that the frame is stored in nano-nodes memory 
	\end{tabular}
	\label{table:parameters}
\end{table*} 

\subsection{Frame storage}
\label{sec:frame_storing}
Once the throughput considering that nano-nodes store a frame for the minimum required time to complete a transmission (i.e., two rounds) is obtained, we extend the analytical model by analyzing the scenario in which nano-nodes are able to keep the frame in memory during an arbitrary $k$ number of rounds. To achieve this, we start by deriving the probability of having a frame stored ($p_{frame}$) as a function of the number of rounds $k$ that nano-nodes keep the frame, with $k\geq 2$. Likewise, we get the expression of not having a frame stored ($p_{empty}$).
To obtain these probabilities we employ the Markov chain shown in Fig. \ref{fig:markov_chain1}, where the state $0$ is associated with not having a frame in memory, and states from $1$ to $k$ stand for having a frame stored in round numbers $1$ to $k$, respectively. 
The transition matrix $T_1$ associated to this Markov chain is defined as:
\begin{equation}
\label{eq:transition_matrix1}
\setlength\arraycolsep{3pt}
T_1= \begin{pmatrix} 
1-p_{rx} & p_{rx} & 0 & 0 & \ldots & 0 & 0 \\
0 & 0 & 1 & 0 & \ldots & 0 & 0  \\
0 & 0 & 0 & 1 & \ldots & 0 & 0 \\
\vdots & \vdots & \vdots & \vdots & \ddots & \vdots & \vdots \\
0 & 0 & 0 & 0 & \ldots & 0 & 1 \\
1-p_{rx} & p_{rx} & 0 & 0 & ... & 0 & 0 \\
\end{pmatrix}_{\!\!(k+1) \times (k+1)}
\end{equation}

As the stationary distribution of a Markov chain with a transition matrix $T$ can be obtained by a vector, $\pi$, so that $\pi T = \pi$, we define the vector $\pi = \{\pi_0, \pi_1, \pi_2, ..., \pi_{k-1}, \pi_k\}$ as the stationary probabilities of being in the states $0$, $1$, $2$, ..., $k-1$, $k$, respectively. Using this property, the series of equations of our Markov chain is defined as follows:
\begin{equation}
\label{eq:markov_equation1}
\pi_0 = (1-p_{rx})\pi_0 + (1-p_{rx})\pi_k
\end{equation}
\begin{equation}
\label{eq:markov_equation2}
\pi_1 = p_{rx}\pi_0 + p_{rx}\pi_k
\end{equation}
\begin{equation}
\label{eq:markov_equation5}
\pi_i = \pi_{i-1}, i = 2, \ldots , k
\end{equation}
\begin{equation}
\label{eq:markov_equation6}
\sum^{k}_{i=0}\pi_i = 1 
\end{equation}

\startfigure
\includegraphics[width=0.48\textwidth]{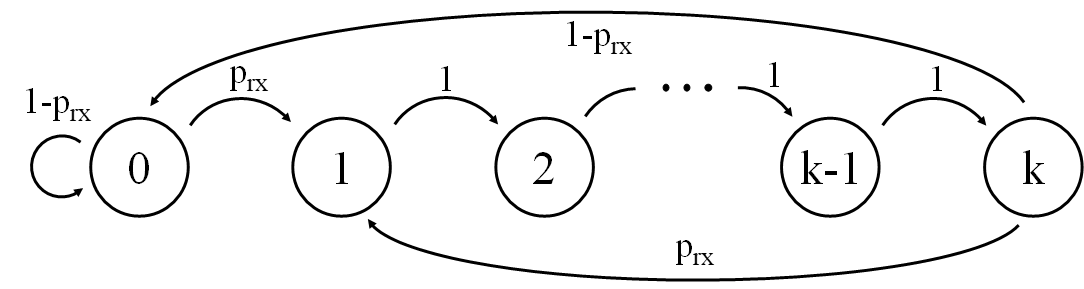}
\caption{Markov chain to derive the probability of having a frame stored. Probabilities of state changes are shown over each transition.}\label{fig:markov_chain1}
\end{figure}

Eq. \ref{eq:markov_equation6} is derived from the second condition of the stationary probabilities; that is, the sum of all the probabilities contained in vector $\pi$ must be 1.

Working with Eqs. \ref{eq:markov_equation1} to \ref{eq:markov_equation6}, we get the following expressions for the values in vector $\pi$:
\begin{equation}
\label{eq:prob_values1}
\pi_0 = \frac{1-p_{rx}}{1+p_{rx}(k-1)}
\end{equation}
\begin{equation}
\begin{gathered}
\label{eq:prob_values2}
\pi_1 = \frac{p_{rx}}{1+p_{rx}(k-1)} \\
\pi_1 = \pi_2 = \ldots = \pi_{k-1} = \pi_k
\end{gathered}
\end{equation}

Once the values in vector $\pi$ have been obtained, we can directly calculate the values of $p_{empty}$ and $p_{frame}$:
\begin{equation}
\begin{gathered}
\label{eq:prob_values3}
p_{empty} = \frac{1-p_{rx}}{1+p_{rx}(k-1)} \\
p_{frame} = \sum^{k}_{i=1}\pi_i = \frac{kp_{rx}}{1+p_{rx}(k-1)}
\end{gathered}
\end{equation}

Upon the expressions of $p_{empty}$ and $p_{frame}$, the raw throughput of the nano-network as a function of the number of rounds that the frame is stored in memory can be calculated following the same procedure as in Eq. \ref{eq:throughput}:
\begin{equation}
\label{eq:throughput_k}
Th_{raw}=n f p_{tx} p_{frame} \left(1-p_{cx}p_{frame}\right)^{n-1}
\end{equation}

As raw throughput considers all the frames received, some of them could be repeated if a nano-node sends the frame stored more than once until it receives a new one. Thus, we also provide an expression for effective throughput (i.e., not considering repeated frames). To do this, we focus on the states of the Markov chain in Fig. \ref{fig:markov_chain1}. As we can see, if the received frame should not be transmitted more than once, there are two valid options: (i) the frame is transmitted in state $1$ or (ii) the frame is transmitted in any of the remaining states (from $2$ to $k-1$) but not in any of the preceding ones (e.g., if the frame is transmitted in state $3$, it cannot have been transmitted in states $1$ and $2$). Note that it cannot be transmitted in state $k$, since the nano-node discards the frame to be ready to receive a new one. Mathematically, considering that only one transmission can be performed per round, we express the probability of transmitting a frame only once during $k$ rounds ($p_{eff}$) as:
\begin{equation}
\begin{gathered}
\label{eq:prob_only1tx}
\begin{split}
\!\!p_{eff} & = \pi_1 p_s + \pi_2 p_s (1-p_s) + \pi_3 p_s (1-p_s)^2 + \ldots \\ 
		& + \pi_{k-1} p_s (1-p_s)^{k-2} = \pi_1 p_s \sum^{k-1}_{i=1}(1-p_s)^{i-1} = \\
		& = \pi_1 \frac{(1-p_s)^k+p_s-1}{p_s-1}	
\end{split}
\end{gathered}
\end{equation}
\noindent where:
	\begin{equation}  
	\label{eq:ps}
	p_{s}=p_{tx}(1-p_{frame}p_{cx})^{n-1}
	\end{equation}

Using Eqs. \ref{eq:prob_values2} and \ref{eq:prob_only1tx}, the effective throughput ($Th_{eff}$) can be directly obtained as:
\begin{equation}
\label{eq:throughput_k_eff}
Th_{eff}= n f \frac{\eta\left((1-p_s)^k+p_s-1\right)}{\left(1+\eta(k-1)\right)\left(p_s-1 \right)}
\end{equation}

\subsection{Quality of Delivery (Q\MakeLowercase{o}D)}
Apart from the throughput, another important metric that can be useful to predict and measure the performance of a THz flow-guided nano-network is the probability of capturing and transmitting a frame before a certain number of rounds. We define this concept as the Quality of Delivery ($QoD$). It will determine how efficiently the nano-network is able to generate and transmit useful data. Thus, employing this metric, the number of nano-nodes required in a nano-network to satisfy a given performance can be more accurately predicted.
To this end, we again rely on Markov chains, specifically the one shown in Fig. \ref{fig:markov_chain2}. For the design of this Markov chain, the following assumptions were made:
\begin{itemize}
	\item State transitions take place in every round, i.e., the state in the Markov chain does not change until a round is totally completed. To achieve this, we define the probability of one nano-node successfully transmitting a frame (without collisions) per round when this frame has been previously received and stored in memory ($p_{s,rnd}$) as:
	\begin{equation}
	\label{eq:ps_round}
	p_{s,rnd} = T f p_{tx} \left(1-p_{cx}p_{frame}\right)^{n-1}
	\end{equation}
	\item Each nano-node stores the frame in memory during a generic number of rounds $k$, following the same approach developed in the previous subsection.
	\item State $Q$ represents the transmission of the stored frame.
	\item Nano-nodes cannot receive and transmit in the same round. We assume a worst-case scenario in which nano-nodes first capture a frame and transmit it, at the earliest, in the following round. Therefore, if a nano-node stores a frame for $k$ rounds, it has $k-1$ chances (one per round) to transmit the frame. This is the reason why state $k$ is not linked to state $Q$, as the round $k+1$ is devoted to receiving a new frame and not to transmitting the former one. 
	

\end{itemize}
The transition matrix $T_{2}$ of this Markov chain is:
	\begin{equation}
	\small
	\label{eq:transition_matrix2}
	\setlength\arraycolsep{2pt}
	\!T_{2}\!= \!\begin{pmatrix} 
		1-p_{rx} & p_{rx} & 0 &  \ldots & 0 & 0 & 0\\[1pt] 
		0 & 0 & 1-p_{s,rnd} & \ldots & 0 & 0 & p_{s,rnd} \\[1pt]
		0 & 0 & 0 & \ldots & 0 & 0 & p_{s,rnd}\\[1pt] 
		\vdots & \vdots & \vdots & \ddots & \vdots & \vdots & \vdots\\[1pt]
		0 & 0 & 0 & \ldots & 0 & 1-p_{s,rnd} & p_{s,rnd}\\[1pt] 
		1-p_{rx} & p_{rx} & 0 & \ldots & 0 & 0 & 0\\[1pt]
		0 & 0 & 0 & \ldots & 0 & 0 & 1\\[1pt]
	\end{pmatrix} 
	\end{equation}
with a size of $(k+2) \times (k+2)$.

Employing this approach, the $QoD$ for one node matches the probability of entering in the state $Q$ after a certain number of rounds $m$, denoted by $\pi_{mQ}$. Thus, we define the vector $\pi_{m} = \{\pi_{m0}, \pi_{m1}, \pi_{m2}, \ldots, \pi_{mk}, \pi_{mQ}\}$ as the probabilities of being in states $0$, $1$, $2$, ..., $k$, $Q$, respectively, after $m$ state transitions. In a Markov chain, $\pi_{m}$ can be calculated as: 
\begin{equation}
	\label{eq:equation_QoS1}
	\pi_{m} = \pi_{0}T_2^m
\end{equation}
where $\pi_{0} = \{\pi_{00}, \pi_{01}, \pi_{02}, ..., \pi_{0k}, \pi_{0Q}\}$ stands for the initial distribution probabilities across the state space of the chain. To get $QoD$ we assume that the nano-network is already working in a stationary state. So, values in vector $\pi_{0}$ equal those in vector $\pi$, obtained in the previous section. The remaining value, $\pi_{0Q}$, is equal to 0.
Thus, from Eq. \ref{eq:equation_QoS1}, we get the expression for every single value of vector $\pi_{m}$ in terms of $k$ and $m$:
\begin{equation}
	\label{eq:equation_QoS2}
	\pi_{mQ} = p_{s,rnd}\sum^{k-1}_{i=1}(\pi_{(m-1)i}) +\pi_{(m-1)Q}	
\end{equation}
\begin{equation}
\label{eq:equation_QoS3}
	\pi_{m0} = (1-p_{rx})\pi_{(m-1)0}+(1-p_{rx})\pi_{(m-1)k}	
\end{equation}
\begin{equation}
\label{eq:equation_QoS4}
	\pi_{m1} =  p_{rx} \pi_{(m-1)0} + p_{rx} \pi_{(m-1)k}
\end{equation}
\begin{equation}
\label{eq:equation_QoS5}
	\pi_{mi} =  (1-p_{s,rnd}) \pi_{(m-1)(i-1)} , i=2,3,\ldots,k		
\end{equation}

Then, once the probability of sending a frame in $m$ rounds for each node has been obtained, we derive the general expression of $QoD$ as follows:
\begin{equation}
\label{eq:equation_QoS}
QoD = 1-(1-\pi_{mQ})^n	
\end{equation}
where $n$ represents the number of nodes in the nano-network. We note that, even though there is not a direct expression for $QoD$, recursive expressions from Eq. \ref{eq:equation_QoS2} to \ref{eq:equation_QoS5} can be easily implemented in any numerical computing environment to obtain the $QoD$ of any flow-guided nano-network, as analyzed in Section \ref{sec:results}.

\startfigure
\includegraphics[width=0.48\textwidth]{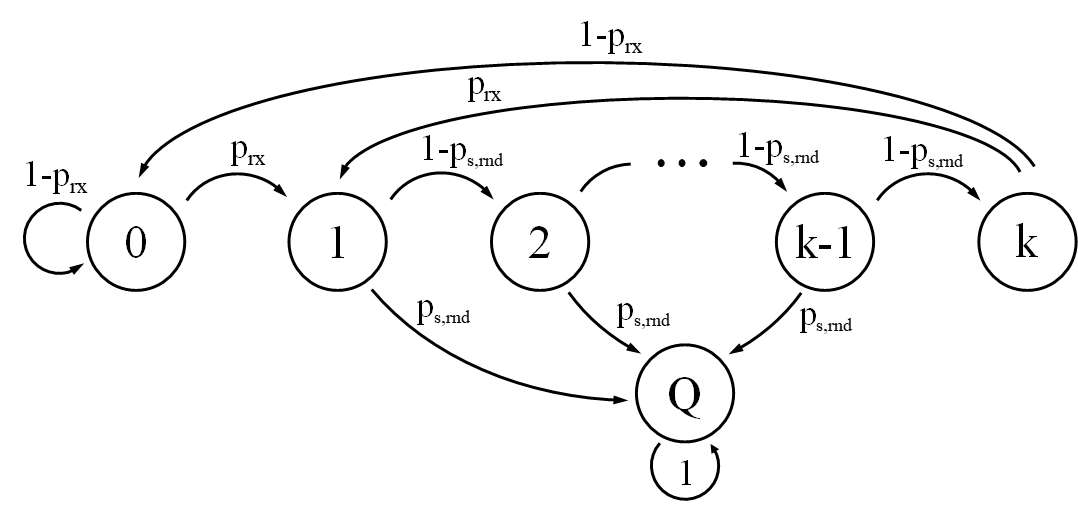}
\caption{Markov chain to obtain $QoD$. Probabilities of state changes are shown over each transition.}\label{fig:markov_chain2}
\end{figure}
 
\subsection{Delay}
Finally, we define the delay of a transmitted frame ($\tau$) as the time elapsed from its reception by a nano-node to the transmission to the nano-router, with $kT \geq \tau \geq 2T$. To provide a consistent metric to measure the average information delay of a nano-network, in this subsection, we derive an analytical expression for the average nano-network delay ($\tau_{av}$). This indicates how fresh the information that the entire nano-network provides to an external agent is, which could be critical for applications that require real-time information streams.

Assuming the same conditions as in the previous section, we define a new Markov chain with a state space comprising $2k$ states, as shown in Fig. \ref{fig:markov_chain3}, with the transition matrix $T_d$ shown in Eq. \ref{eq:transition_matrix3}.

%
\begin{figure*}
	\setcounter{MaxMatrixCols}{15}
	\begin{equation}
	\label{eq:transition_matrix3}
	\begingroup
	\setlength\arraycolsep{4pt}
	T_{d}= \left(\begin{matrix}
	1-p_{rx} & \: p_{rx} & 0 & \dotsm & \: 0 & 0 & 0 & 0 & \dotsm & \: 0 & 0\\
	0 & 0 & 1-p_{s,rnd} & \dotsm & \: 0 & 0 & p_{s,rnd} & 0 &\dotsm & \: 0 & 0\\
	0 & 0 & 0 & \dotsm & \: 0 & 0 & 0 & p_{s,rnd} &\dotsm & \: 0 & 0\\
	\vdots & \vdots & \vdots & \ddots & \: \vdots & \vdots & \vdots & \vdots &\ddots & \: \vdots & \vdots\\
	0 & 0 & 0 & \dotsm & \: 0 & \: 1-p_{s,rnd} & 0 & 0 &\dotsm & \: 0 & \: p_{s,rnd}\\
	1-p_{rx} & p_{rx} & 0 & \dotsm & \: 0 & 0 & 0 & 0 &\dotsm & \: 0 & 0\\
	0 & 0 & 0 & \dotsm & \: 0 & 0 & 0 & 1 & \dotsm & \: 0 & 0\\
	0 & 0 & 0 & \dotsm & \: 0 & 0 & 0 & 0 & \dotsm & \: 0 & 0\\
	\vdots & \vdots & \vdots&\ddots & \: \vdots & \vdots & \vdots & \vdots &\ddots & \: \vdots & \vdots\\
	0 & 0 & 0 & \dotsm & \: 0 & 0 & 0 & 0 &\dotsm & \: 0 & 1\\
	1-p_{rx} & p_{rx} & 0 & \dotsm  & \: 0 & 0 & 0 & 0 &\dotsm & \: 0 & 0\\
	\end{matrix}\right)_{2k \times 2k}
	\endgroup
	\end{equation}
\end{figure*}
As in Subsection \ref{sec:frame_storing}, we rely on the stationary distribution of a Markov chain to derive the expression of the average delay of a transmitted frame. Thus, we define the vector $\pi^d = \{\pi^d_0, \pi^d_1, \pi^d_2, \ldots, \pi^d_k, \pi^d_{Q2},\pi^d_{Q3}, \ldots, \pi^d_{Qk}\}$, so that $\pi^dT_d=\pi^d$. The first $k+1$ states, from $0$ to $k$, are devoted to nano-node states in which the frame is stored but it has not been transmitted to the nano-router, as in the previous subsection. States from $Q_2$ to $Q_{k}$ represent the fact that the frame has been transmitted in a specific round, from $2$ to $k$ (we should note that the frame cannot be transmitted in the same round as that in which it has been received; that is, the first round). As a transmitted frame should be counted only once (if the same frame is transmitted twice, the second transmission is discarded), when one of the $Q$ states is reached, the next state transitions will be in the remaining $Q$ states until round number $k$. Based on the expression $\pi^dT_d=\pi^d$, the equations associated with the stationary probabilities are:
\begin{equation}
\label{eq:delay_equation1}
\pi^d_0 = (1-p_{rx})\pi^d_0 + (1-p_{rx})\pi^d_k+(1-p_{rx})\pi^d_{Qk}
\end{equation}
\begin{equation}
\label{eq:delay_equation2}
\pi^d_1 = p_{rx}\pi^d_0 + p_{rx}\pi^d_k + p_{rx}\pi^d_{Qk}
\end{equation}
\begin{equation}
\label{eq:delay_equation3}
\pi^d_i = (1-p_{s,rnd})^{i-1}\pi^d_1, i = 2,3,\ldots,k
\end{equation}
\begin{equation}
\label{eq:delay_equation4}
\pi^d_{Qi} = p_{s,rnd}\sum^{i}_{j=2}(1-p_{s,rnd})^{j-2}\pi^d_1, i = 2,3,\ldots,k
\end{equation}
\begin{equation}
\label{eq:delay_equation5}
\sum^{k}_{i=0}{\pi^d_{i}}+\sum^{k}_{i=2}{\pi^d_{Qi}} = 1
\end{equation}

Working with this series of equations, values in vector $\pi^d$ are defined as follows:
\begin{equation}
\label{eq:delay_values1}
\pi^d_0 = \frac{1-p_{rx}}{1+p_{rx}(k-1)}
\end{equation}
\begin{equation}
\label{eq:delay_values2}
\pi^d_i =\frac{p_{rx}(1-p_{s,rnd})^{i-1}}{1+(k-1)p_{rx}}, i = 1,2,\ldots,k
\end{equation}
\begin{equation}
\label{eq:delay_values3}
\pi^d_{Qi}\! = \!\frac{p_{rx}((1-p_{s,rnd})^{i}\!+\!p_{s,rnd}\!-\!1)}{(p_{s,rnd}-1)((k-1)p_{rx}+1)}, i \!=\! 2,3,\ldots,k
\end{equation}

According to the design of the Markov chain, the general expression of the average delay for the frames transmitted can be calculated as:
\begin{equation}
\label{eq:delay}
	\tau_{av} = \sum^k_{i=2} \frac{\pi^d_{Qi}-\pi^d_{Q(i-1)}}{\pi^d_{Qk}}\cdot iT
\end{equation}
where $T$ stands for the time required to complete a round. We note that the minimum average delay in this expression is $2$$\cdot$$T$, since we assume that a nano-node takes at least two complete rounds to send a frame.
\begin{figure}
\centering
\includegraphics[width=0.48\textwidth]{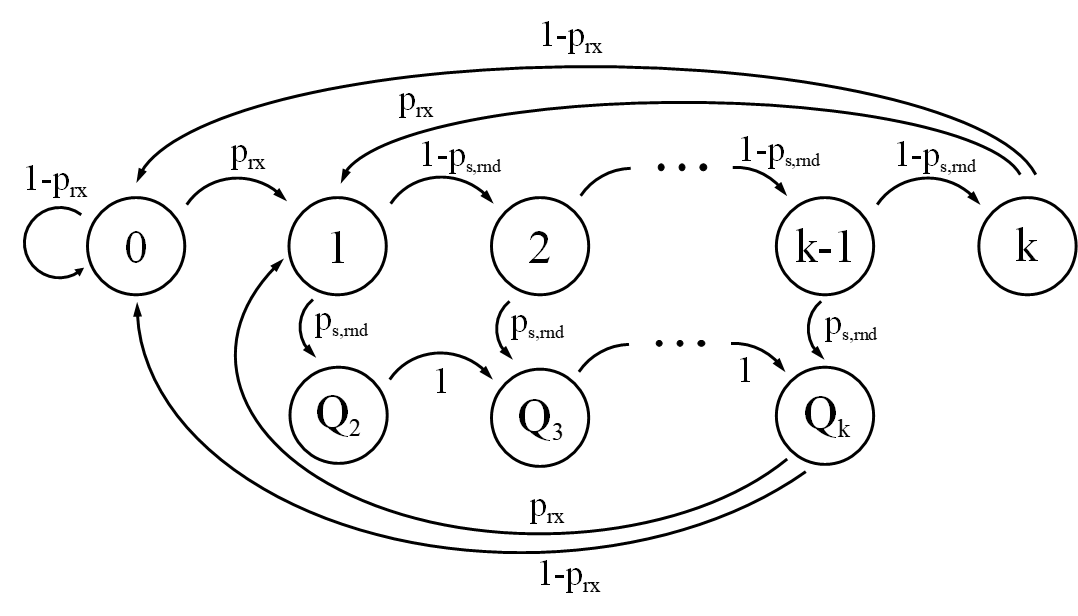}
\caption{Markov chain to derive the average delay. Probabilities of state changes are shown over each transition.}\label{fig:markov_chain3}
\end{figure}
\section{Results} 
\label{sec:results}

In this section, we first evaluate network performance (considering the figures of merit defined in Section \ref{sec:analytical_model}) as a function of different parameters, so that the influence of each one of them on the performance of a THz flow-guided nano-network is thoroughly discussed. Then, we study the feasibility of using flow-guided nano-networks in medical applications in terms of their network performance. To do this, we apply the analytical model to the applications described in Section \ref{sec:applications}, employing realistic and accepted values for all the parameters employed. All the results obtained throughout this section have also been validated by means of simulations. Simulations have been implemented in MATLAB with the following scenario variables: number of nano-nodes, communication range, diameter of vein, battery charging frequency, nano-node velocities in the flow, time to complete one round through the system, time to transmit a complete frame, total volume of the flow, and simulation seed. The nano-network has been simulated considering a branched, closed-loop medium (modeling the cardiovascular system) where nano-nodes flow for an equivalent (simulated) time of 1 hour. Nano-nodes flow through each branch of the circuit, simulating veins, with a certain probability equal to the actual percentage of blood flowing through that vein, containing a total volume of 5 liters (total volume of blood in the human body) \cite{circulatorysystem}. The average time to complete a round through the cardiovascular system ($T$) is set to 60 seconds \cite{circulatorysystem}.
Every point in the simulations has been repeated ten times with a different seed.

\subsection{Performance evaluation of a generic flow-guided nano-network}
In this subsection, we conduct a sensitivity analysis for the four most relevant parameters: (i) number of nano-nodes, (ii) diameter of vein, (iii) communication range, and (iv) number of rounds storing a frame. 
To this end, we set the values of the remaining parameters as follows:
\begin{itemize}
	\item The time to transmit a complete frame ($t_f$) is determined by two factors: (i) the frame length (in bits) and (ii) the symbol rate. Regarding the first factor, we assume 64 bits, so frames are short enough to be stored and transmitted by nano-nodes but large enough to include accurate medical measurement and some fields of an ad-hoc protocol, aimed at preserving the integrity and robustness of the transmitted data. For the second factor, we assume a conservative nano-node symbol rate of $10^6$ symbols per second, in accordance with the related literature \cite{Jornet2014,Singh2018,Bourgeois2018,10.1145/3345312.3345465}. Thus, assuming an \textit{On–Off keying} modulation \cite{Jornet2014} where one symbol equals one bit, the bitrate ($R$) is equal to $1$ Mbps and the value of $t_f$ is 64 $\mu$s. 
	\item As nano-nodes are powered by means of a piezoelectric generator, we assume that the mechanical strength caused by a heart rate of 60 beats per minute will allow us to harvest enough energy to activate and transmit or receive a frame once per second; that is, $1/f$ = 1 s. To justify this claim, we analyze the energy balance in each nano-node once a frame has been received, as data transmission is the most energy-consuming task. On the one hand, the energy expected to be consumed in each cycle is given by:
	\begin{equation}
	E_{node} = L_{f} W E_{p} + \frac{L_{f} P_{bit}}{f}
	\end{equation}
	where $W$ is the probability of transmitting a pulse, since the transmission of a logical "1" is carried out by sending a pulse, whereas a logical "0" is sent as silence. To ensure that the nano-network is working in all cases, we assume $W$ = 1 to cover the worst-case scenario, that is, all the frames are composed of "1"s (the energy consumption is at its maximum). $E_p$ stands for the energy of a pulse, which is considered 0.1 fJ to get a communication range of 1 mm \cite{Canovas-Carrasco2019}. This range was calculated considering the particular characteristics of THz waves propagation inside the human body, specifically in blood, and a transmission power of 1 mW using the TS-OOK modulation.
	$L_f$ is the length of the frame (64 bits) and $P_{bit}$ denotes the power required by the circuitry to keep a bit stored in memory (2.4 fW \cite{6585815}).  
	On the other hand, we consider that each nano-node features a nano-generator able to gather $\Delta Q$ = 6 pC per compress-release cycle (induced by each heartbeat) with a generated voltage ($V_g$) of 0.2 V, as analyzed in the literature \cite{Canovas-Carrasco2019,Canovas-Carrasco2018,Jornet2012b}, and a nano-capacitor with a capacitance ($C$) of 10 pF --with a maximum energy capacity $\left(E_{max} = \frac{C V_g^2}{2}\right)$ of 200 fJ--. Using these parameters, the energy harvesting rate ($\lambda_h$) can be calculated by this expression (taken from \cite{Canovas-Carrasco2019}):
	\begin{equation}
	\lambda_{h}(E_{nc}) = \Delta Q f_{ng} V_g \sqrt{\frac{E_{nc}}{E_{max}}} \left(1\!-\!\sqrt{\frac{E_{nc}}{E_{max}}}\right)
	\end{equation}
	where $E_{nc}$ is the energy already stored in the nano-capacitor and $f_{ng}$ is the frequency of compress-release cycles (1 Hz). Using this expression and applying the energy consumption of the nano-node with $W$ = 1, the nano-generator is able to perform a frame communication every second and stabilize the energy level at around 140 fJ. 
	
	\item The velocity of nano-nodes when passing through the nano-router coverage zone ($v$) is set in accordance with the velocity of blood flow in a superficial vein in the hand \cite{Klarhofer2001}; that is, 10 cm/s on average \cite{Cephalic}. 
	\item The percentage of the blood flow ($\eta$) flowing through the bio-sensor is set to 10\%, taking as a reference a renal vein \cite{renalvein} in which a bio-sensor could be implanted to detect bacteria from kidneys.
	\item The diameter of the vein where the nano-router is placed ($D$) is set to 6 mm, provided that it is not the parameter under study. This value is in line with a generic application in which the nano-router is placed in a superficial vein in the hand \cite{Cephalic}.
	\item When applicable, the communication range ($r$) is established at 1 mm, a worst-case value for the THz-based communication in blood according to both the high path-loss in this medium and the low energy employed to generate the EM pulses of the modulation in use \cite{Canovas-Carrasco2019}.
\end{itemize}


\subsubsection{\textbf{Number of nano-nodes}}
\begin{figure*}
	\setlength{\tabcolsep}{-1pt}
	\begin{tabular}{c c c}
		\begin{subfigure}[b]{0.34\textwidth}
			\centering
			\includegraphics[width=1.0\linewidth]{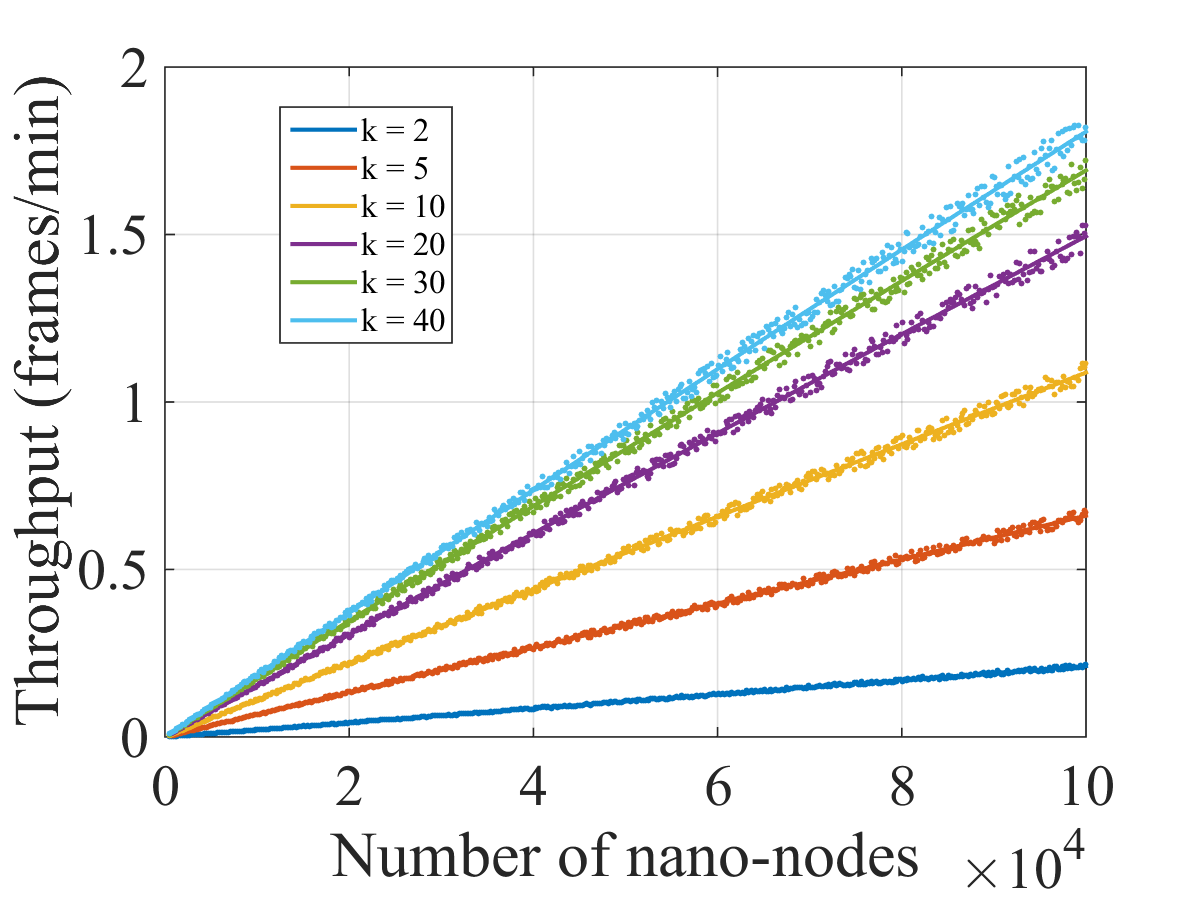}
			\label{fig:throughput1}
		\end{subfigure}
		&
		\begin{subfigure}[b]{0.34\textwidth}
			\centering
			\includegraphics[width=1.0\linewidth]{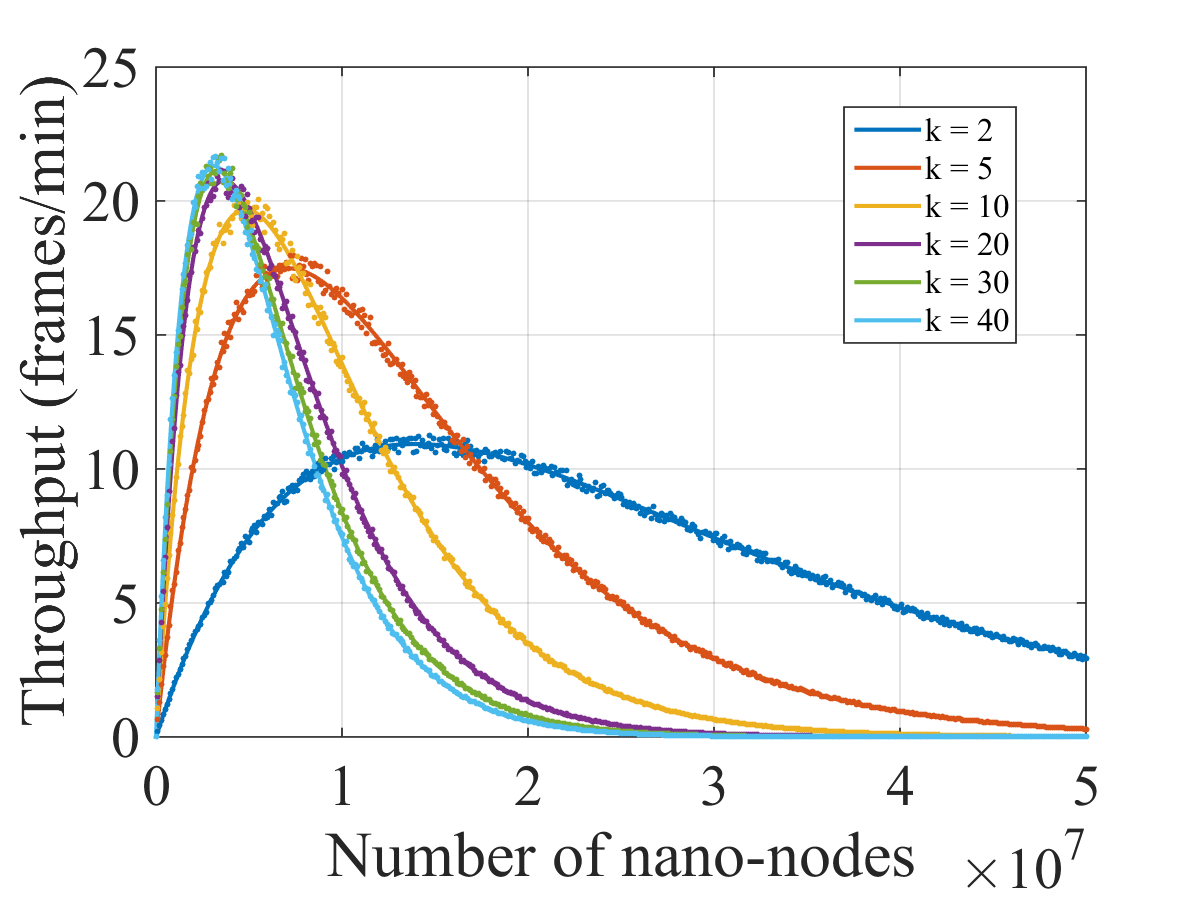}
			\label{fig:throughput2}
		\end{subfigure}
		&
		\begin{subfigure}[b]{0.34\textwidth}
		\centering
		\includegraphics[width=1.0\linewidth]{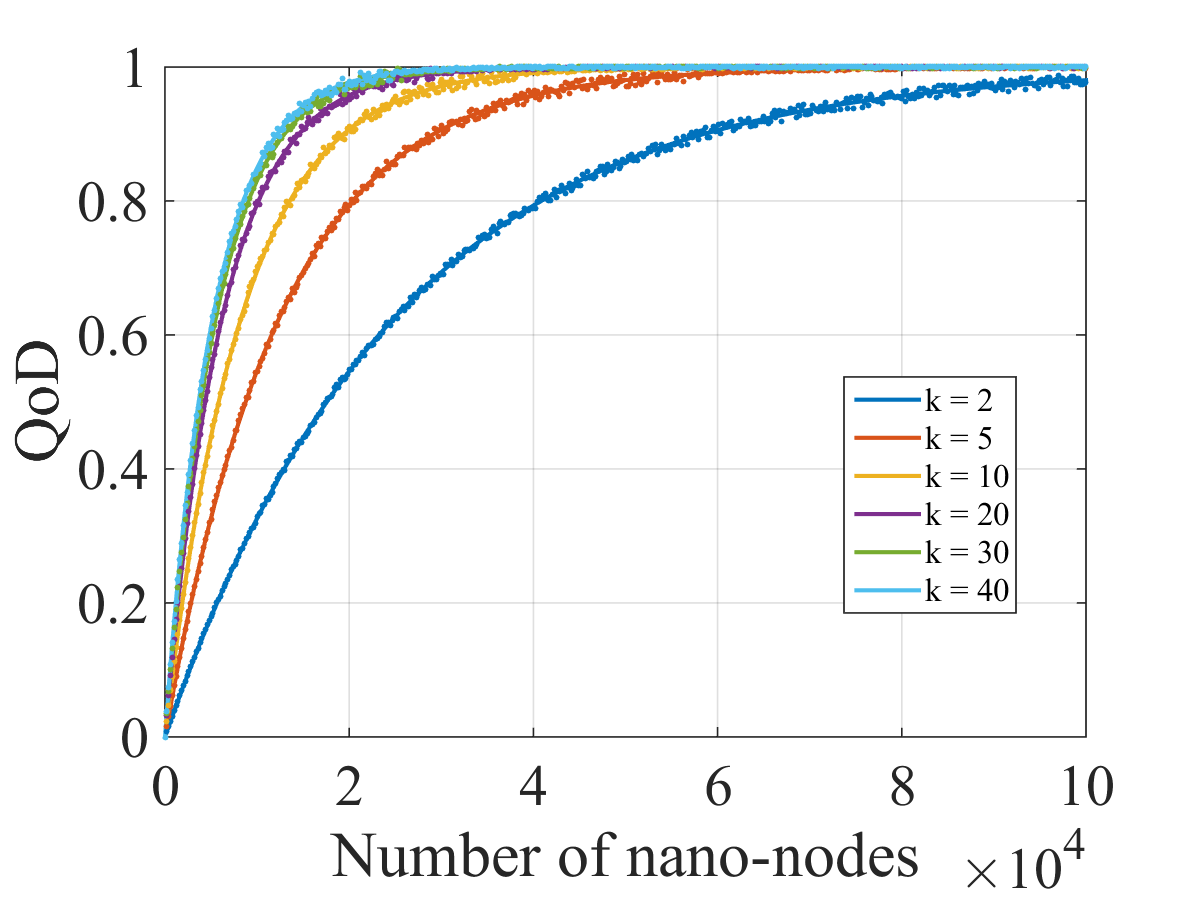}
		\label{fig:QoD_nodes}
	\end{subfigure}\\
	\end{tabular}
	\caption{Nano-network throughput (a), (b) and $QoD$ (c) versus number of nano-nodes for different values of $k$. Lines show the throughput obtained from the proposed analytical model and dots from simulations.}
	\label{fig:throughput}
\end{figure*}

Using all the parameters above, Fig. \ref{fig:throughput} represents the throughput and $QoD$ as a function of the number of nano-nodes in the nano-network for different values of $k$, both theoretically (solid line) and obtained by simulation (dots). 

As can be seen in Fig. \ref{fig:throughput}a, nano-network throughput increases almost linearly as the number of nodes grows with all the values of $k$, since the number of collisions is still negligible. The impact of $k$ on the throughput can also be observed, as it increases the probability of a nano-node having one frame stored in memory. However, when the number of nano-nodes is high enough, collisions become relevant (see Fig. \ref{fig:throughput}b) and the throughput reaches a maximum. Then, it progressively decreases until it reaches zero due to the increasing number of collisions. 
Another notable fact is that when $k$ increases (that is, the percentage of nano-nodes ready to transmit a frame is higher), the maximum throughput is greater. This occurs, fundamentally, because in order to get this value the density of nodes in the circuit is less than with lower values of $k$, thus reducing the number of collisions for a similar amount of transmitting nano-nodes. It can also be noted that as $k$ increases, the enhancement of the throughput lessens, since the percentage of nano-nodes with a frame stored in memory increases to 100\%.

The same trend can be observed in the $QoD$ selected metric. Concretely, Fig. \ref{fig:throughput}c shows the $QoD$ of a flow-guided nano-network with $m$ = 10 (that is, the probability of transmitting a frame before 10 rounds) as a function of the number of nano-nodes. As $n$ increases, the value of $QoD$ gets closer to one. Also in this case, the effect of $k$ is noticeable, showing a significantly higher $QoD$ as $k$ grows but moving toward a maximum value when it becomes excessively high. 

\begin{figure*}
\centering
	\begin{tabular}{c c}
		\begin{subfigure}[b]{0.4\textwidth}
			\centering
			\includegraphics[width=1\textwidth]{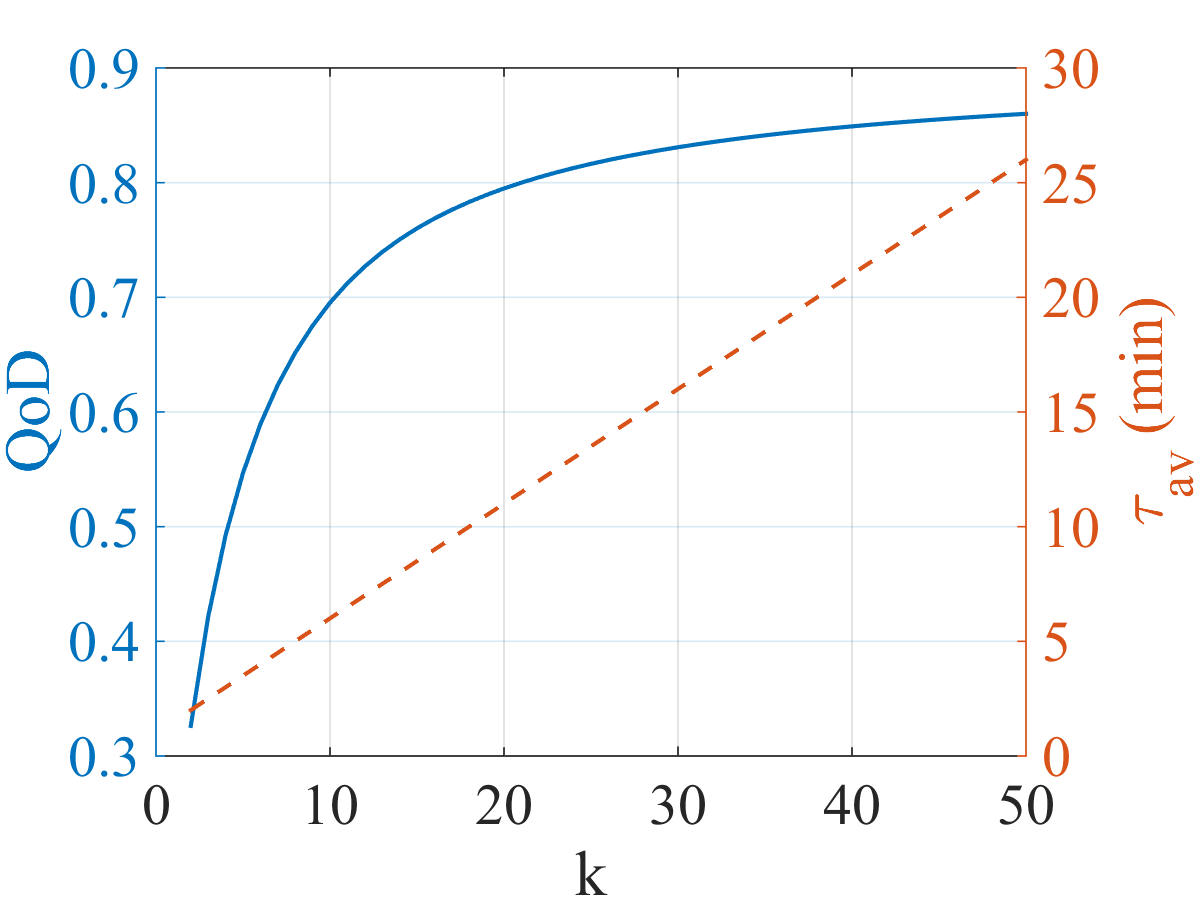}
			\caption{}
			\label{fig:QoS_delay}
		\end{subfigure}
		&
		\begin{subfigure}[b]{0.4\textwidth}
			\centering
			\includegraphics[width=1\textwidth]{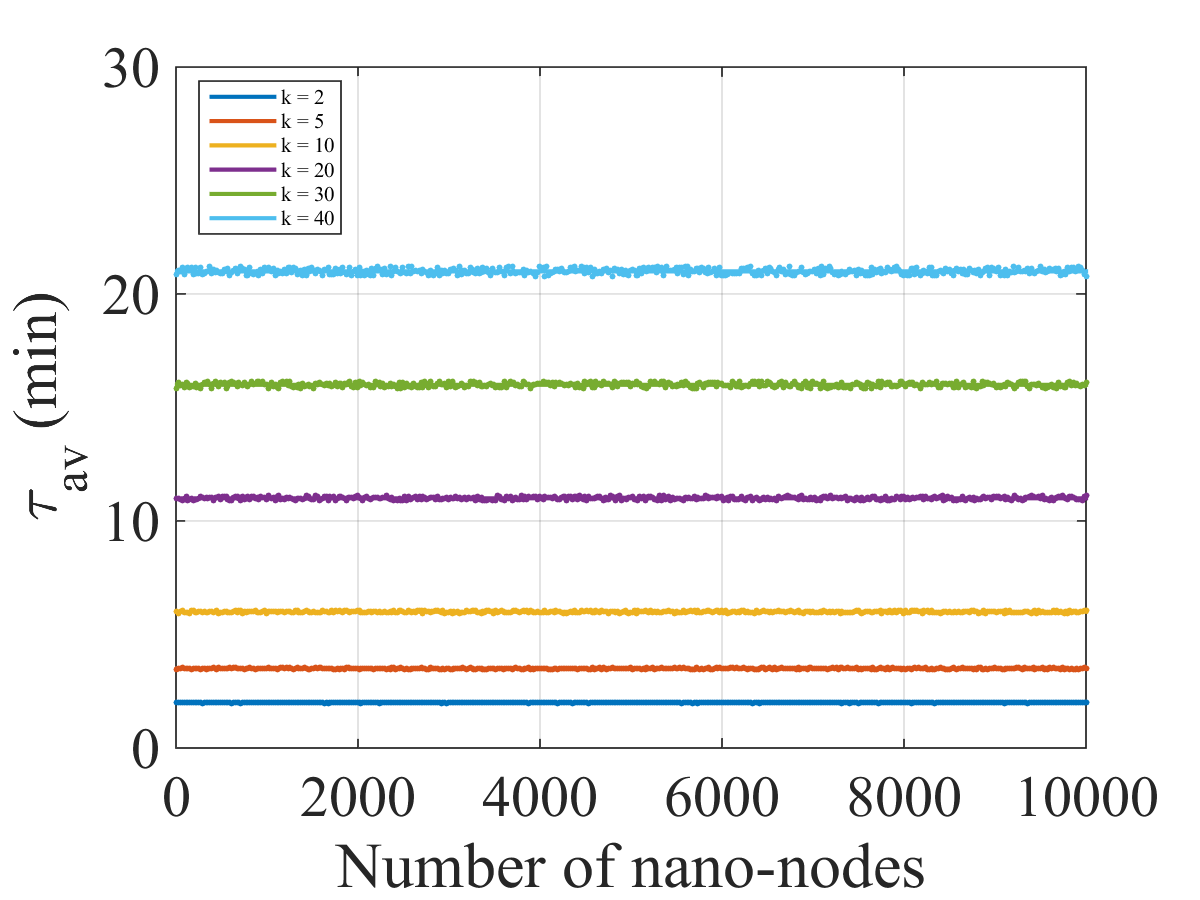}
			\caption{}
			\label{fig:delay}
		\end{subfigure}\\
	\end{tabular}
	\caption{(a) $QoD$ and average data delay ($\tau_{av}$) versus $k$ with $n$ = $10^4$ and $m$ = 10. (b) $\tau_{av}$ versus number of nano-nodes for different values of $k$. Lines show the throughput obtained from the proposed analytical model and dots from simulations.}
	\label{fig:QoS_and_tau}
\end{figure*}

So far, raising the value of $k$ has had a positive impact on nano-network performance in terms of throughput and $QoD$. However, as shown in Fig. \ref{fig:QoS_delay}, there is a significant drawback to increasing $k$ because this increases the average delay ($\tau_{av}$) of the transmitted frames. When nano-nodes keep a frame in memory for a longer time, the availability is higher, entailing a higher global throughput, but the freshness of these frames is compromised. It is also worth noting that, in this scenario, the average delay of the transmitted frames is completely independent of the number of nano-devices in the network (see Fig. \ref{fig:delay}). From these results, we can conclude that the time each nano-node needs, on average, to transmit a frame from its reception does not change, although the number of nodes that can transmit becomes higher. Therefore, there is a trade-off between throughput and information freshness when varying $k$ that should be considered in the design of a flow-guided nano-network. 

%
%

\subsubsection{\textbf{Diameter of vein}}
\begin{figure*}
	\setlength{\tabcolsep}{-1pt}
	\begin{tabular}{ccc}
		\begin{subfigure}[b]{0.34\textwidth}
			\centering
			\includegraphics[width=1.0\linewidth]{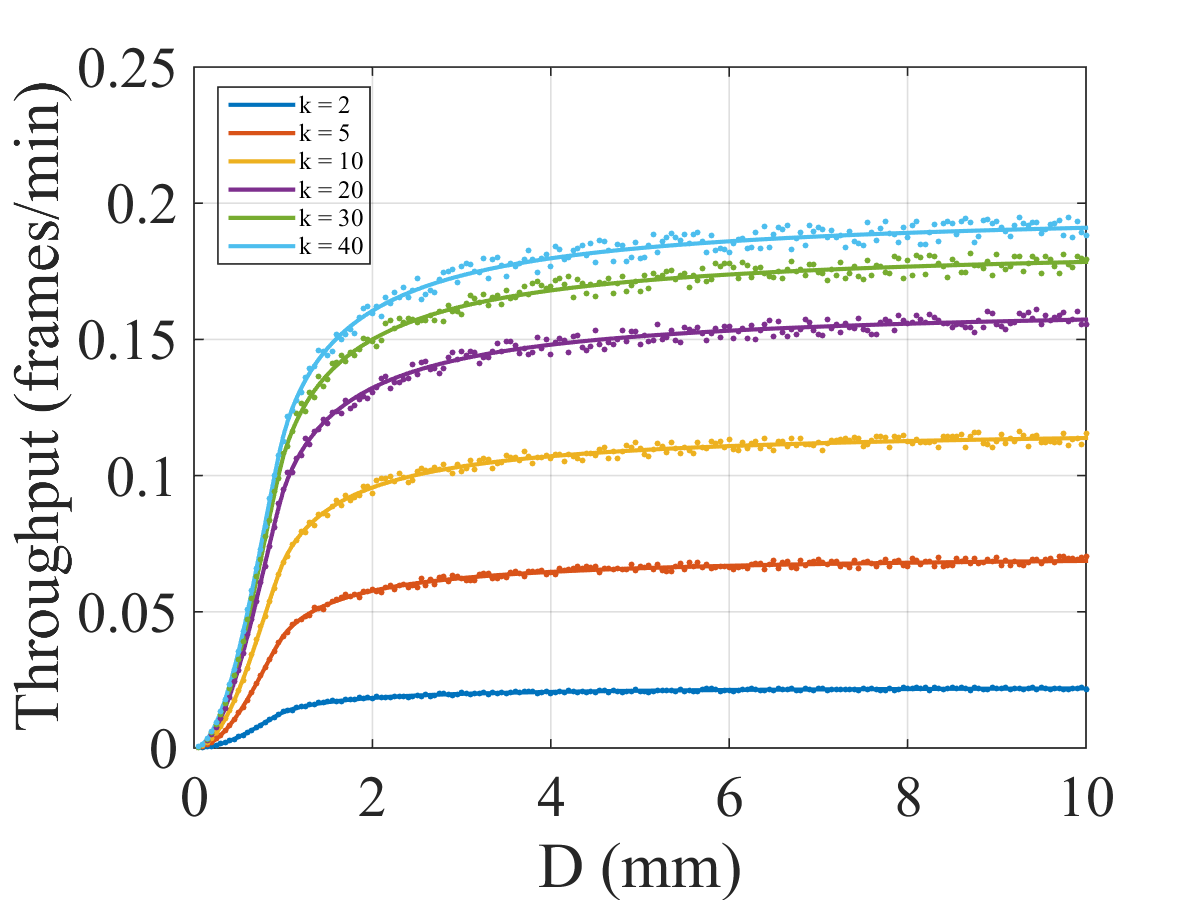}
			\caption{}
			\label{fig:th_D1}
		\end{subfigure}
		&
		\begin{subfigure}[b]{0.34\textwidth}
			\centering
			\includegraphics[width=1.0\linewidth]{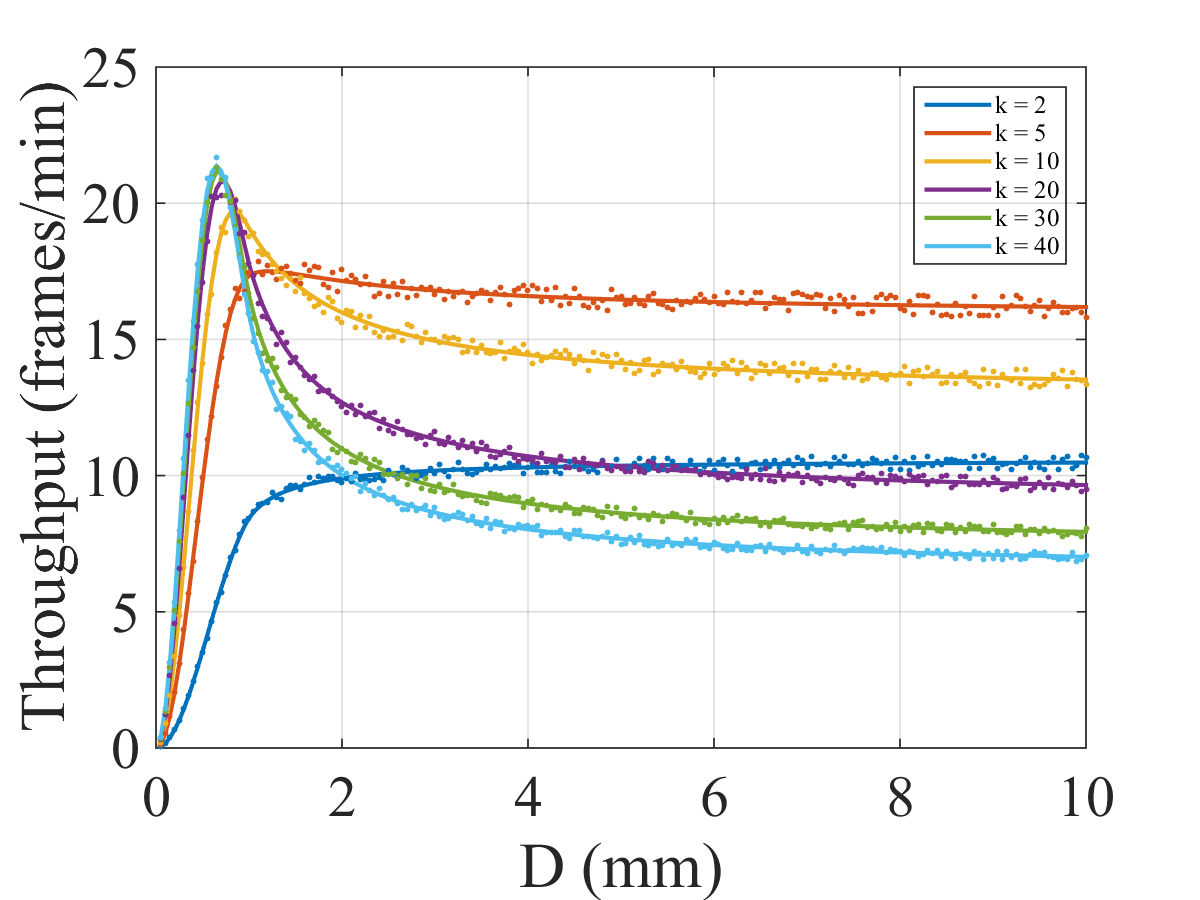}
			\caption{}
			\label{fig:th_D2}
		\end{subfigure}
		&
		\begin{subfigure}[b]{0.34\textwidth}
			\centering
			\includegraphics[width=1.0\linewidth]{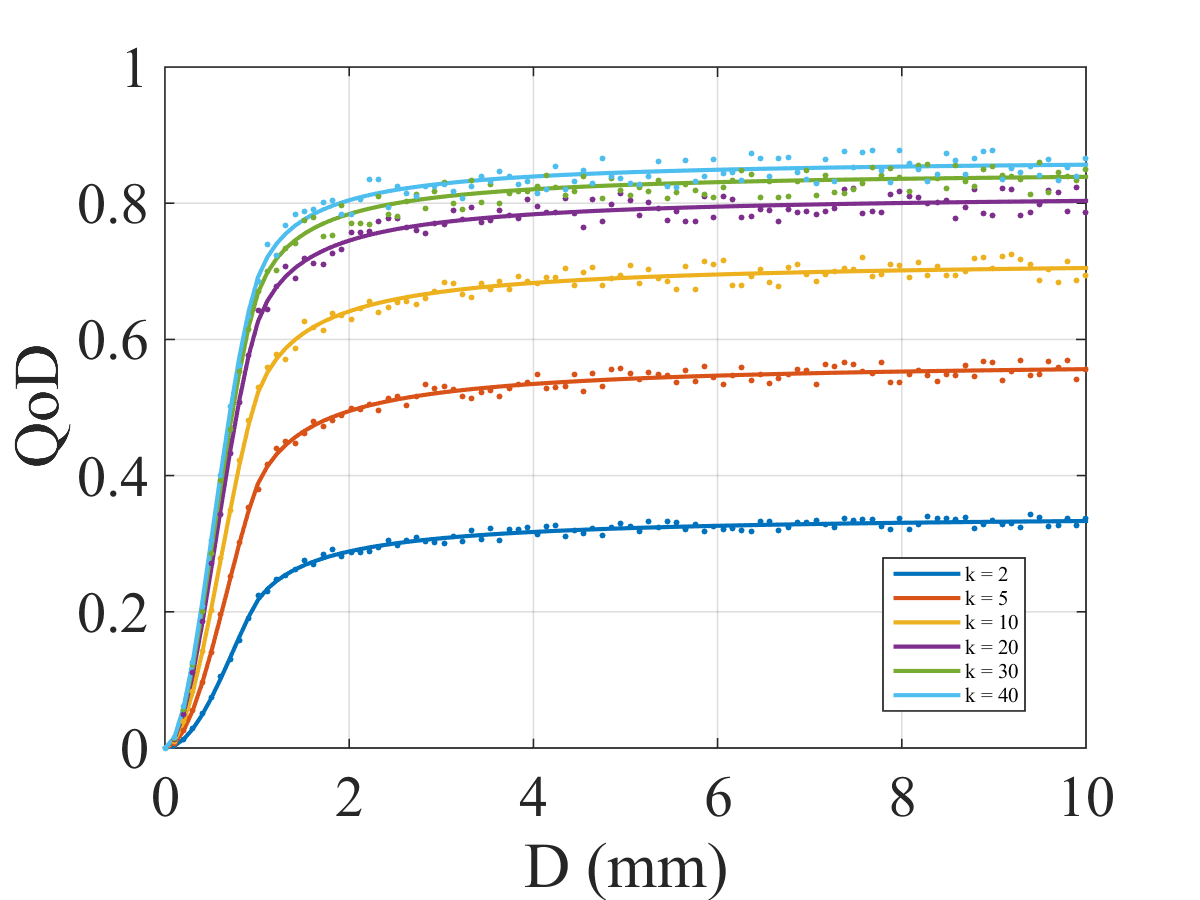}
			\caption{}
			\label{fig:QoD_D}
		\end{subfigure}\\
	
	\end{tabular}
	\caption{Nano-network throughput and $QoD$ versus $D$ for different values of $k$. In (a), $n$ = $10^4$. In (b), $n$ = $10^7$, In (c), $n$ = $10^4$ and $m$ = 10. Lines show the throughput obtained from the proposed analytical model and dots from simulations.}\label{fig:th_D}
\end{figure*}

Another relevant factor in nano-network performance is the diameter of the vein in which the nano-router is implanted. Fig. \ref{fig:th_D} shows the nano-network throughput as a function of $D$ for two different values of $n$. As can be noted in Fig \ref{fig:th_D1}, if $n$ is not excessively high (i.e., collisions are not relevant yet), the throughput noticeably grows for values of $D$ below 2 mm and then it stabilizes for larger diameters. This result is derived from the communication range of nano-devices, set to 1 mm, so the throughput grows as $V_{cv}$ gets larger. Interestingly, if the nano-network is working in a state in which collisions are not numerous, we can confirm that placing the nano-router in thicker veins does not penalize nano-network performance.
However, if $n$ grows and collisions become more frequent, there is an optimal $D$ where the throughput is maximized (see. Fig \ref{fig:th_D2}) at around 0.75 mm. This is due to the fact that transmission and collision zones increase at different paces as a function of $D$. The collision zone increases faster than the transmission zone, making the impact of collisions higher for larger diameters. Therefore, to reach the maximum throughput in a scenario where the number of collisions is substantial, the value of $D$ should be optimized. The $QoD$ shows similar behavior for $10^4$ nano-nodes, as shown in Fig. \ref{fig:QoD_D}, with a sharp increase for low diameters and stabilizing when $D$ grows. This is because the number of collisions for $10^4$ nano-nodes is not relevant. If $n$ increases, the value of $QoD$ tends to be 1, reducing the impact of $D$, since the number of nano-nodes needed to hinder the value of $QoD$ is too high (higher than $10^7$) to be relevant.


\subsubsection{\textbf{Communication range}}
\begin{figure}
\vspace*{-\baselineskip}
	\centering
	\includegraphics[width=0.4\textwidth]{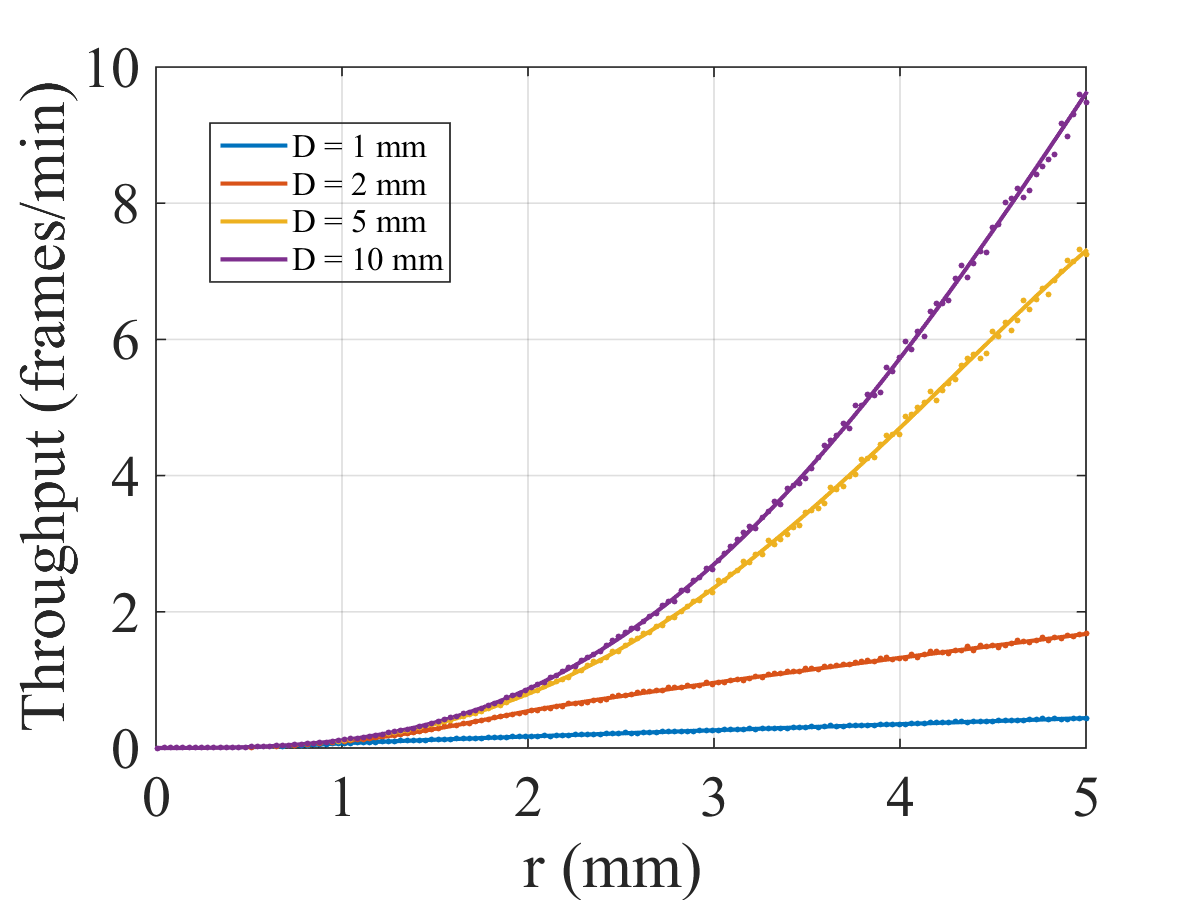}
	\caption{Nano-network throughput versus communication range ($r$) with $n$ = $10^4$ and $k$ = 10. Lines show the throughput obtained from the proposed analytical model and dots from simulations.}\label{fig:th_r}
\end{figure}

Even though the high path loss in watery mediums strictly limits the communication range of nano-devices, we should also analyze its impact on nano-network performance and highlight its relevance, even for small variations. The fact that future research could reveal that there are some parts of the human body where blood flow presents lower absorption at THz frequencies (extending the communication range of nano-devices), therefore making this analysis useful to highlight how performance can be affected. Thereby, Fig. \ref{fig:th_r} shows the throughput as a function of the communication range ($r$), varying it from 0 to 5 mm (values in line with the ranges expected for nano-networks \cite{Akyildiz2010,Canovas-Carrasco2018a}), with $10^4$ nano-nodes. As can be noted, the impact of $r$ on nano-network performance is the most significant among all the parameters analyzed. For example, the act of increasing $r$ from 1 to 2 mm implies an improvement of approximately 8 times in throughput for $D$ = 5 mm and $k$ = 10. The larger $D$ is, the more noticeable the enhancement will be, since the volume of the vein within the coverage range is greater. If progress in nano-communications enabled nano-devices to increase $r$ up to 5 mm inside the human body, nano-network throughput would increase by a factor of approximately 70, drastically boosting the potential of flow-guided nano-networks.

\subsection{Application}
\label{sec:results_app}
In this subsection, we analyze potential flow-guided nano-networks based on the applications described in Section \ref{sec:applications}. For all the applications, the nano-router would be located on the \textit{cephalic} vein, a superficial vein in the hand that would allow communication between the nano-router and an external device, as studied in \cite{Canovas-Carrasco2018a}. On average, the diameter of this vein is about 6 mm with a flow velocity of 10.9 cm/s, in accordance with the study in \cite{Cephalic}. Regarding the communication range of nano-nodes, we consider a worst-case scenario in which the communication range is limited to 1 mm. All the parameters considered in this case of study are specified in Table \ref{table:parameters_app}.
 
\begin{figure}
	\vspace*{-\baselineskip}
	\centering
	\includegraphics[width=0.4\textwidth]{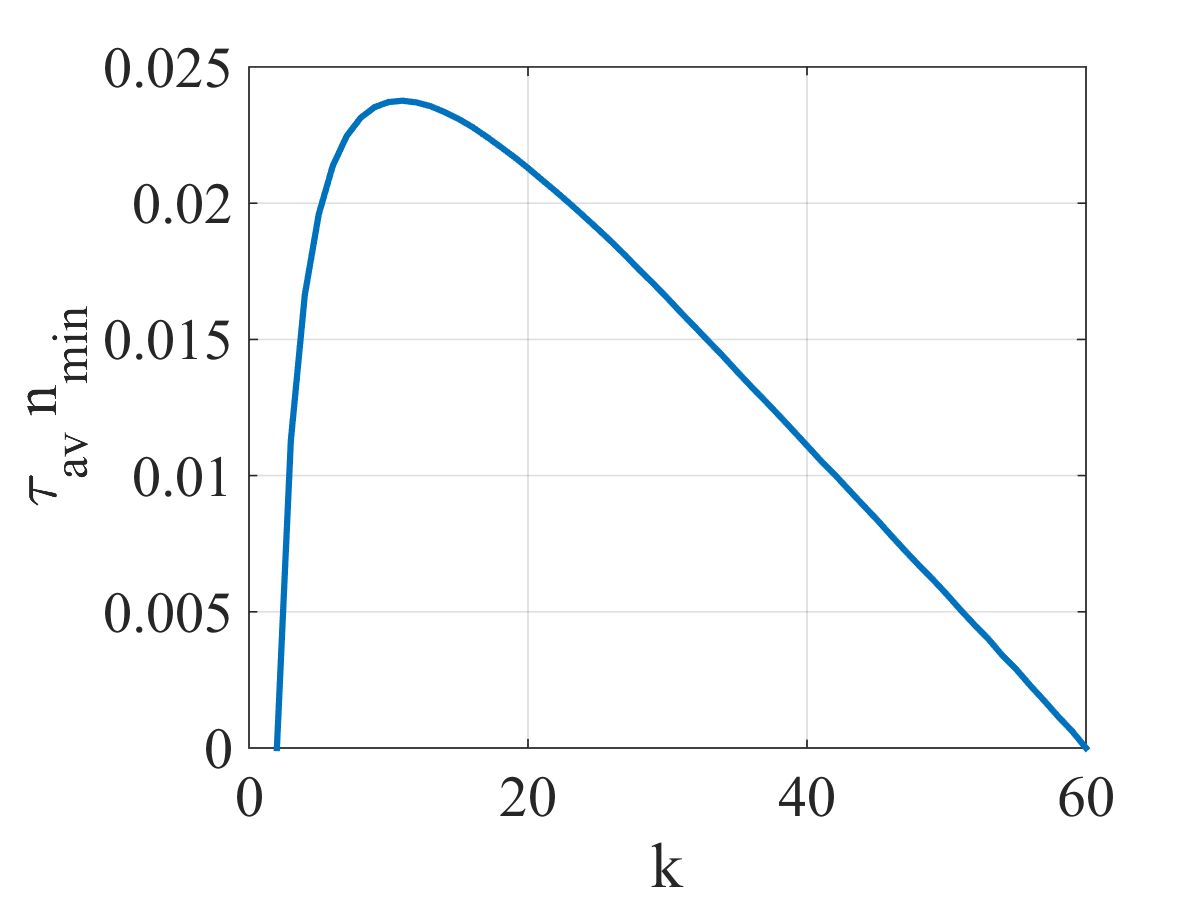}
	\vspace*{-2 mm}
	\caption{$\tau_{av} \cdot n_{min}$ vs $k$, so that the value of $QoD$ is equal to 0.99	for $m$ = $m_{target}$.}\label{fig:tau_n_k}
\end{figure}

\begin{table}[]
	\centering
	\renewcommand{\arraystretch}{1.2}
	\caption{Parameters employed in Section \ref{sec:results_app}.}
	\begin{tabular}{c c} 
		\hline
		Parameter & Value \\
		\hline
		$D$ & 6 mm\\
		$v$ & 10.9 cm/s\\		
		$r$ & 1 mm\\
		$R$ & 1 Mbps\\ 
		$L_f$ & 64 bits\\
		$C$ & 10 pF\\
		$V_g$ & 0.2 V\\
		$\Delta Q$ & 6 pC\\
		$E_p$  & 0.1 fJ\\
		$P_{bit}$ & 2.4 fW\\
		$1/f$ & 1 s\\
	\end{tabular}
	\vspace*{-\baselineskip}
	\label{table:parameters_app}
\end{table} 
For the first application (bacterial blood infections), it is necessary to locate a bio-sensor in a renal vein ($\eta$ = 0.1 \cite{renalvein}) which is able to send a warning frame (to the nano-router) before 1 hour (3600 seconds) from the moment the level of bacteria reaches a given threshold. In order to ensure this condition with a probability of 99\%, three parameters must be taken into account: $QoD$, $\tau_{av}$ and $k$. On one hand, the number of nano-nodes in the nano-network should be high enough to ensure a $QoD$ equal to 0.99 for a certain number of rounds ($m_{target}$), defined as: $m_{target}$ = $ \lfloor \tau_{target}/T \rfloor$, where $\tau_{target}$ is the target deadline required for the application under study. In this case, $\tau_{target}$ = 3600 s, with the value of $m_{target}$ being equal to 60 rounds. On the other hand, $QoD$ also depends on the value of $k$, with $m_{target} \geq k \geq 2$, obtaining higher values as $k$ increases (see Fig. \ref{fig:throughput}c). However, as shown in Fig. \ref{fig:QoS_delay}, when $k$ grows, the freshness of the transmitted frames is compromised, implying an increase of $\tau_{av}$. Therefore, to calculate the values of $n$ and $k$ to meet a specific deadline, a trade-off between $QoD$ and $\tau_{av}$ should be considered to achieve a good nano-network performance with acceptable information freshness. To this end, we define the metric $\tau_{av} \cdot n_{min}$, with $n_{min}$ as the minimum number of nano-nodes necessary to obtain a $QoD$ equal to 0.99. This metric provides a balance between the number of nodes required (that decreases when $k$ gets higher) and the average frame delay (that increases with $k$). Fig. \ref{fig:tau_n_k} shows the normalized value of this metric as a function of $k$. The maximum value of $\tau_{av} \cdot n_{min}$ indicates the equilibrium point between $n_{min}$ and $\tau_{av}$, obtaining the optimal value of $k$. When $k$ is excessively low, $n_{min}$ gets too high, even though $\tau_{av}$ is at a minimum, resulting in a metric value close to zero. Similarly, when $k$ becomes too high, $n_{min}$ decreases but the value of $\tau_{av}$ becomes too high to meet the application requirement.
Results reveal that a nano-network with 6246 nano-nodes storing a frame for 11 rounds ($k$ = 11) should be able to send a frame before 1 hour (i.e., 60 rounds) with a $QoD$ of 0.99, thus being enough to provide robust and reliable service. The throughput provided by this nano-network is 0.073 frames/min (i.e., one frame every 13.7 minutes, on average), with $\tau_{av}$ equal to 6.5 minutes.	

The following three applications (i.e., viral load monitoring, sepsis, and heart attacks) have similar requirements since the nano-network must be designed to send a warning before a specific time limit. For the viral load monitoring application, the bio-sensor should be located in the cephalic vein ($\eta$ = 0.0056 \cite{Cephalic}) and the warning must be delivered before 1 day ($\tau_{target}$ = 86400 s). Following the same procedure as in the previous application, the optimal value of $k$ to minimize the number of nano-nodes is 55, while the nano-network would need 580 nano-nodes to meet the requirements.
Regarding sepsis detection, the bio-sensor is placed in the jugular vein ($\eta$ = 0.14 \cite{Jugular1}), the deadline is 1 hour ($\tau_{target}$ = 3600 s), while for heart attacks the bio-sensor is positioned in the cava vein ($\eta$ = 0.35 \cite{cavavein}), and the warning must be reported before 15 minutes ($\tau_{target}$ = 900 s). In these two cases, the minimum number of nano-nodes in the nano-network is 5397 ($k$ = 11) and 19294 ($k$ = 6), respectively. As can be observed, the impact of the target deadline on the size of the nano-network is much greater than the location of the bio-sensor.

Finally, the last application (restenosis) requires a periodic update of the state of the bio-sensor, located in a coronary artery ($\eta$ = 0.03 \cite{coronaryartery}), so the performance requirement relies on the throughput and the average delay. An update period of 1 hour should be enough to keep medical staff well informed about a patient's condition. In this case, we size the nano-network to get a throughput of 0.033 frames per minute (2 frames per hour), with $\tau_{av}$ = 30 minutes. Thus, the number nano-nodes required for this application would be 2328. The nano-network features obtained with the proposed analytical model for each application are shown in Table \ref{table:app_results}.

\begin{table*}[]
	\centering
	\renewcommand{\arraystretch}{1.2}
	\caption{Nano-network features for each application.}
	\begin{tabular}{c c c c c c} 
		\hline
		Application & $n$ & $k$ & $\tau_{target}$& Throughput (frames/min) & $\tau_{av}$ (min) \\
		\hline
		Bacterial blood infections & 6246 & 11 & 1 hour & 0.073 & 6.5 \\
		Viral load monitoring  & 580 & 55 & 1 day & 0.003  & 28.5 \\
		Sepsis & 5397 & 11 & 1 hour & 0.074  & 6.5 \\
		Heart attacks & 19294 & 6 & 15 min & 0.286  & 4 \\
		Restenosis  & 2328 & 58 & - & 0.033  & 30 \\
	\end{tabular}
	\label{table:app_results}
\end{table*}

%

\section{Conclusion}
\label{sec:conclusion}

The use of the THz band in the field of nano-communications could be a real breakthrough in applied medicine and revolutionize the way information about the state of human organisms and particular tissues is collected. In pursuit of this objective, in this paper we have conceptually presented an in-body flow-guided nano-network composed of three different types of devices: (i) bio-sensors measuring medical parameters, (ii) nano-nodes circulating within the bloodstream and working as data carriers, and (ii) nano-routers performing as gateways and forwarding vital medical information acquired from the nano-network to external devices directly connected to medical personnel. As thoroughly reviewed, this nano-network would be able to play an essential role in diverse applications, concretely, in the early-stage detection of bacterial and viral infections, sepsis, heart attacks, and restenosis.

Based on this nano-network architecture, we have derived an analytical model jointly considering the peculiarities of nano-devices (i.e., energy balance and communication in the THz band) and the complex branched nature of the cardiovascular system. This model accurately predicts nano-network performance, employing different figures of merit; that is, throughput, Quality of Delivery (QoD), and average transmission delay. In addition, this model can be used to adequately project the number of nano-nodes required for each application.
Nano-network performance has been analyzed as a function of different parameters, highlighting the impact of (i) the number of nano-nodes, (ii) the number of rounds that a frame is stored in nano-nodes memory, (iii) the diameter of the vein where the nano-router is implanted, and (iv) the nano-nodes communication range. All the results obtained have also been validated by means of simulations.
Finally, we have analyze nano-network dimensioning based on the particular requirements of each medical application. To this end, we have tuned the parameter $k$ to minimize the number of nano-nodes employed to meet the specifications. Outcomes reveal that a nano-network employing a reasonable number of nano-nodes would be able to effectively fulfill medical application specifications, reducing intrusiveness. Keeping in mind the ultra-small size of these nano-nodes, the number of devices considered (between 580 and 20000), deployed in-body, should be seen as non-invasive. 

To sum up, this work has shown that THz nano-communications can pave the way for multiple and promising medical applications, being able to perform useful tasks without needing a huge number of nano-devices. Besides, the analytical model proposed here represents a step forward in the design and deployment of flow-guided nano-networks, as it is generic enough to be directly employed or easily adapted to many other different applications, widening the scope of THz flow-guided nano-network applicability. 


\bibliographystyle{IEEEtran}
\bibliography{nano}

\begin{IEEEbiography}[{\includegraphics[width=1in,height=1.25in,clip,keepaspectratio]{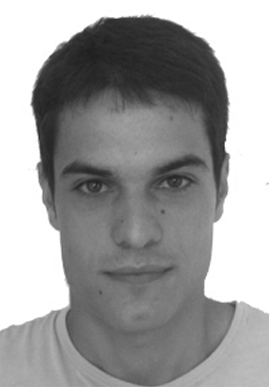}}]{Sebastian Canovas-Carrasco} received the B.S. degree in telecommunication systems engineering and the M.S. degree in telecommunications engineering from the Universidad PolitÈcnica de Cartagena, Spain, in 2014 and 2016, respectively, where he is currently pursuing the Ph.D. degree. His research interest includes the electromagnetic wireless nanonetworks, radiocommunications at the THz band, and machine learning.
\end{IEEEbiography}

\begin{IEEEbiography}[{\includegraphics[width=1in,height=1.25in,clip,keepaspectratio]{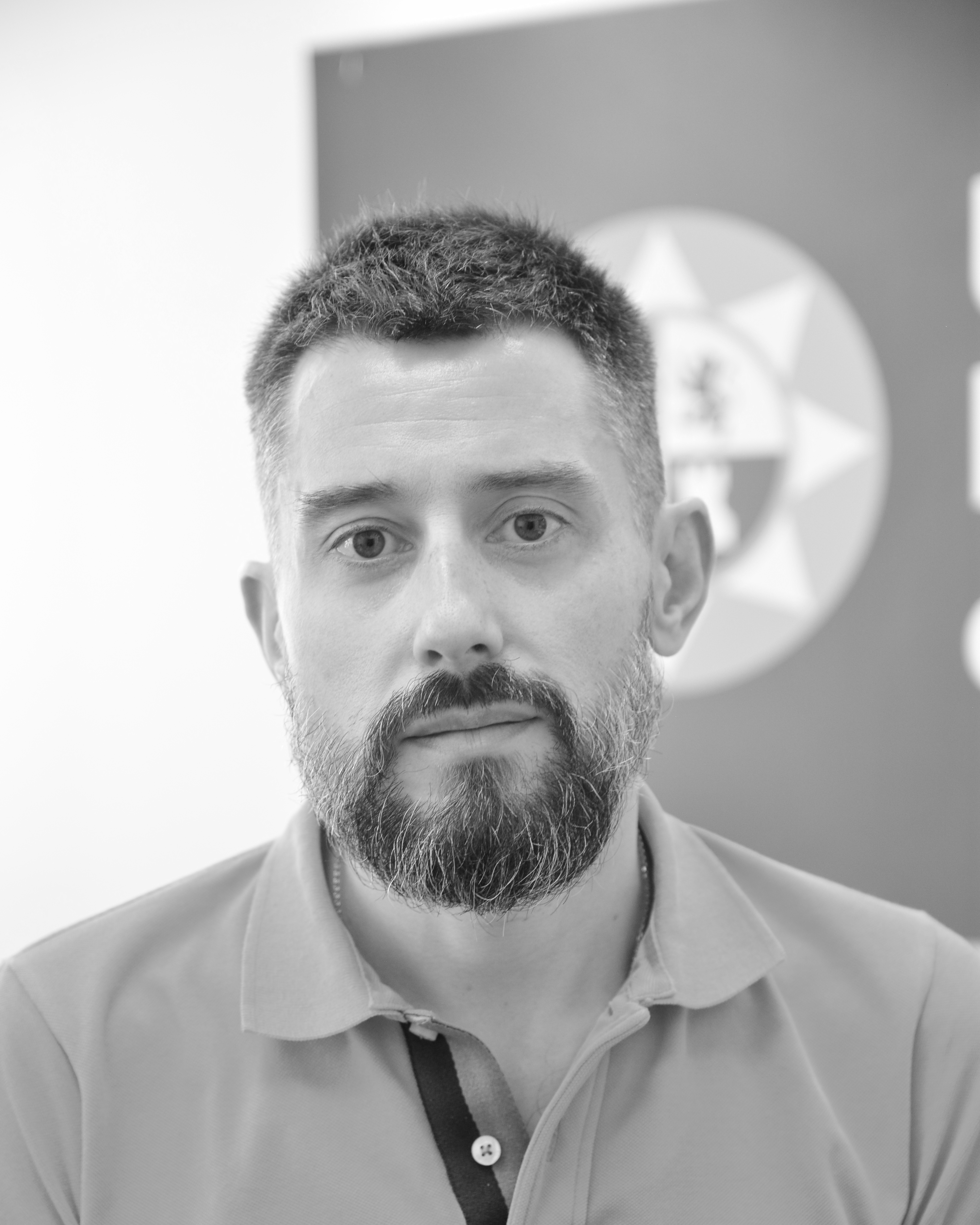}}]{Rafael Asorey-Cacheda} received his M.Sc. degree in Telecommunication Engineering (major in Telematics and Best Master Thesis Award) and his Ph.D. (cum laude and Best PhD Thesis Award) in Telecommunication Engineering from the Universidade de Vigo, Spain, in 2006 and 2009, respectively. He was a researcher with the Information Technologies Group, University of Vigo, Spain until 2009. Between 2008 and 2009 he was also R\&D Manager at Optare Solutions, a Spanish telecommunications company. Between 2009 and 2012, he held an Ángeles Alvariño position, Xunta de Galicia, Spain. Between 2012 and 2018, he was an associate professor at the Centro Universitario de la Defensa en la Escuela Naval Militar, Universidade de Vigo. Currently, he is an associate professor at the Universidad Politécnica de Cartagena. He is author or co-author of more than 60 journal and conference papers, mainly in the fields of switching, wireless networking and content distribution. He has been a visiting scholar at New Mexico State University, USA (2007-2011) and at Universidad Politécnica de Cartagena, Spain (2011, 2015). His interests include content distribution, high-performance switching, peer-to-peer networking, wireless networks, and nano-networks.
\end{IEEEbiography}

\begin{IEEEbiography}[{\includegraphics[width=1in,height=1.25in,clip,keepaspectratio]{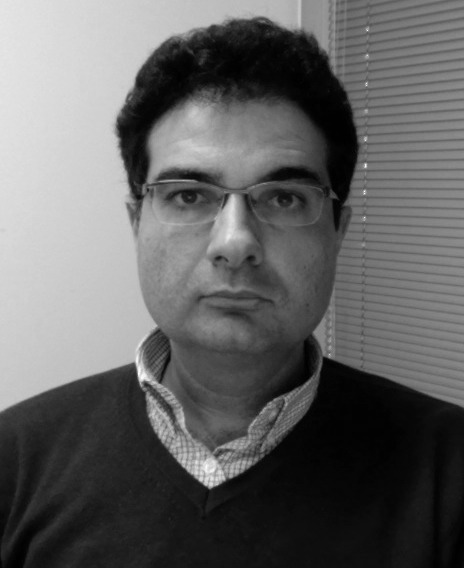}}]{Antonio-Javier Garcia-Sanchez} received his Industrial Engineering degree (M.S. degree) in 2000 from the Universidad Politécnica de Cartagena (UPCT), Spain. Since 2001, he has joined the Department of Information Technologies and Communications (DTIC), UPCT, obtaining his Ph.D. in 2005. Currently, he is an Associate Professor at the UPCT. He is (co)author of more than 70 conference and journal papers, thirty of them indexed in the Journal Citation Report (JCR). He has been the head of several research projects in the field of communication networks and optimization, and he currently is a reviewer of several journals listed in the ISI-JCR. He is also inventor/co-inventor of 9 patents or utility models, and he has been a TPC member or Chair in about thirty International Congresses or Workshops. His main research interests are in the areas of wireless sensor networks (WSNs), streaming services, smart grid, IoT, and nano-networks. 
\end{IEEEbiography}

\begin{IEEEbiography}[{\includegraphics[width=1in,height=1.25in,clip,keepaspectratio]{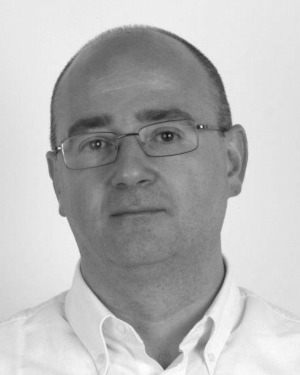}}]{Joan Garcia-Haro} (M’91) received his M.S. degree and Ph.D in telecommunication engineering from the Universitat Politècnica de Catalunya, Barcelona, Spain, in 1989 and 1995, respectively. He is currently a Professor with the Universidad Politécnica de Cartagena (UPCT). He is author or co-author of more than 70 journal papers mainly in the fields of switching, wireless networking and performance evaluation. Prof. Garcia-Haro served as Editor-in-Chief of the IEEE Global Communications Newsletter, included in the IEEE Communications Magazine, from April 2002 to December 2004. He has been Technical Editor of the same magazine from March 2001 to December 2011. He also received an Honorable Mention for the IEEE Communications Society Best Tutorial paper Award (1995). He has been a visiting scholar at Queen’s University at Kingston, Canada (1991-1992) and at Cornell University, Ithaca, USA (2010- 2011).
\end{IEEEbiography}

\begin{IEEEbiography}[{\includegraphics[width=1in,height=1.25in,clip,keepaspectratio]{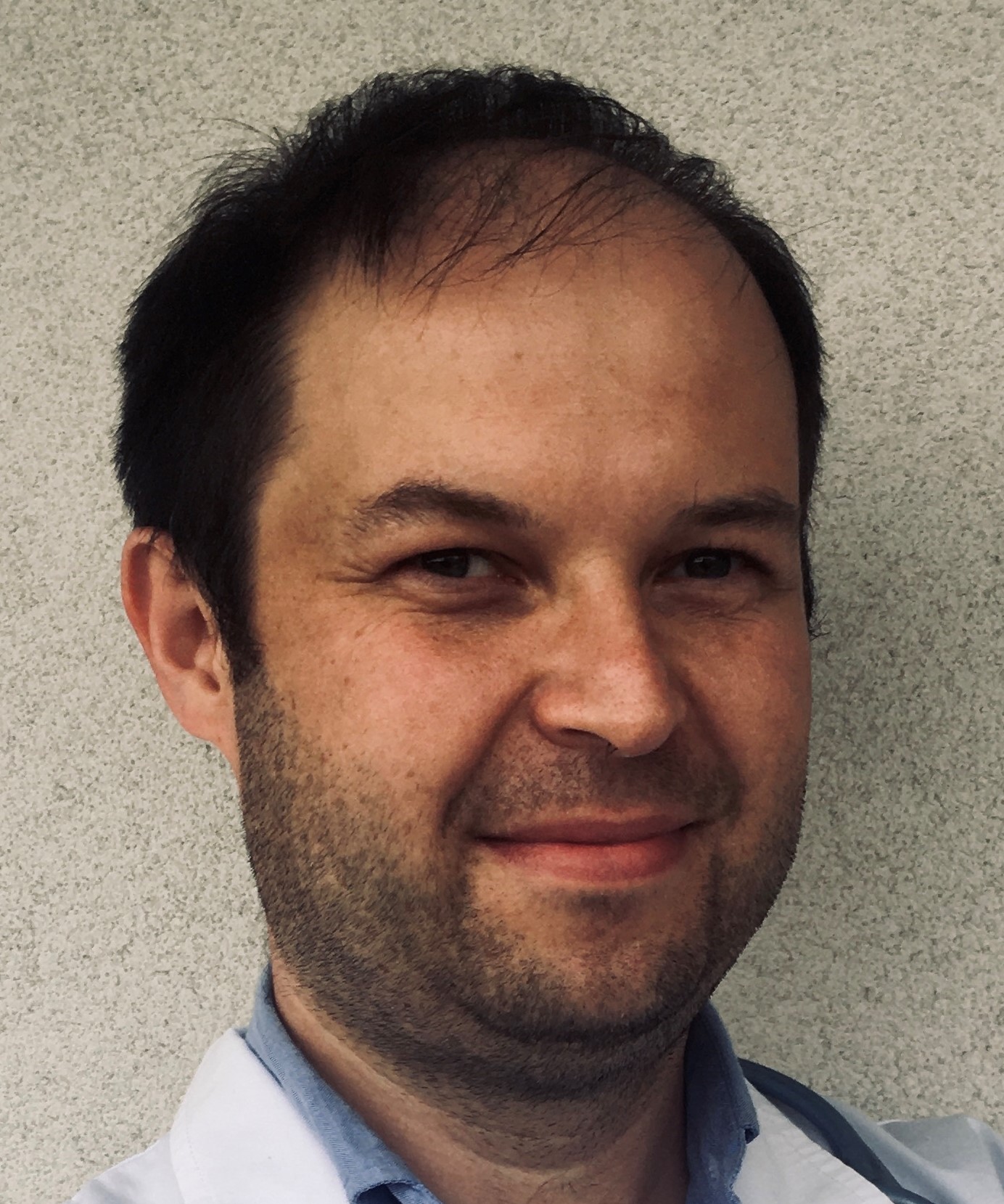}}]{Krzysztof Wojcik} received his M.Sc. and Ph.D. degrees in biophysics from the Jagiellonian University in Krakow, Poland 2003 and 2015, respectively, and an M.D. from the Jagiellonian University Medical College in Krakow, Poland 2007. He was an Assistant at Division of Cell Biophysics Faculty of Biochemistry, Biophysics and Biotechnology Jagiellonian University (2007-2014). At present he is an Assistant Professor Department of Allergy, Autoimmunity and Hypercoagulability in II Chair of Internal Medicine JUMC. His research interests include systemic vasculitis, confocal microscopy and STED (Stimulated Emission Depletion) techniques and their applications in autoantibodies research, as well as the use of fluorescent probes in nanocommunications.
\end{IEEEbiography}

\begin{IEEEbiography}[{\includegraphics[width=1in,height=1.25in,clip,keepaspectratio]{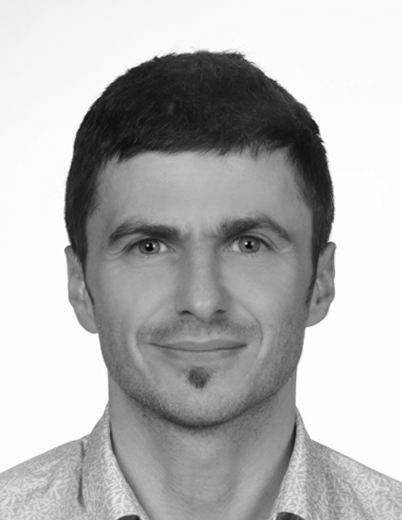}}]{Pawel Kulakowski} received his Ph.D. in telecommunications from the AGH University of Science and Technology in Krakow, Poland, in 2007, and currently he is working there as an assistant professor. He spent about 2 years in total as a post-doc or a visiting professor at Technical University of Cartagena, University of Girona, University of Castilla-La Mancha and University of Seville. He was involved in research projects, especially European COST Actions: COST2100, IC1004 and CA15104 IRACON, focusing on topics of wireless sensor networks, indoor localization and wireless communications in general. His current research interests include molecular communications and nano-networks. He was recognized with several scientific distinctions, including 3 awards for his conference papers and a governmental scholarship for young outstanding researchers.
\end{IEEEbiography}

\EOD

\end{document}